\documentclass[aps,prl,preprint,tightenlines,superscriptaddress,showpacs,byrevtex]{revtex4}
\usepackage{graphicx} 
\usepackage{dcolumn}  

\graphicspath{{ps}}

\begin{document}

  \hfill\begin{minipage}[l]{6cm}{Belle Prerpint 2006-38\\
KEK   Preprint 2006-55}\end{minipage}

\title{ \quad\\[0.5cm]  Study of $\bar{B^0}\to {D}^0 \pi^+ \pi^-$ decays}

\affiliation{Budker Institute of Nuclear Physics, Novosibirsk}
\affiliation{Chiba University, Chiba}
\affiliation{Chonnam National University, Kwangju}
\affiliation{University of Cincinnati, Cincinnati, Ohio 45221}
\affiliation{Department of Physics, Fu Jen Catholic University, Taipei}
\affiliation{The Graduate University for Advanced Studies, Hayama, Japan} 
\affiliation{University of Hawaii, Honolulu, Hawaii 96822}
\affiliation{High Energy Accelerator Research Organization (KEK), Tsukuba}
\affiliation{Hiroshima Institute of Technology, Hiroshima}
\affiliation{Institute of High Energy Physics, Chinese Academy of Sciences, Beijing}
\affiliation{Institute of High Energy Physics, Vienna}
\affiliation{Institute of High Energy Physics, Protvino}
\affiliation{Institute for Theoretical and Experimental Physics, Moscow}
\affiliation{J. Stefan Institute, Ljubljana}
\affiliation{Kanagawa University, Yokohama}
\affiliation{Korea University, Seoul}
\affiliation{Kyungpook National University, Taegu}
\affiliation{Swiss Federal Institute of Technology of Lausanne, EPFL, Lausanne}
\affiliation{University of Ljubljana, Ljubljana}
\affiliation{University of Maribor, Maribor}
\affiliation{University of Melbourne, Victoria}
\affiliation{Nagoya University, Nagoya}
\affiliation{Nara Women's University, Nara}
\affiliation{National Central University, Chung-li}
\affiliation{National United University, Miao Li}
\affiliation{Department of Physics, National Taiwan University, Taipei}
\affiliation{H. Niewodniczanski Institute of Nuclear Physics, Krakow}
\affiliation{Nippon Dental University, Niigata}
\affiliation{Niigata University, Niigata}
\affiliation{University of Nova Gorica, Nova Gorica}
\affiliation{Osaka City University, Osaka}
\affiliation{Osaka University, Osaka}
\affiliation{Panjab University, Chandigarh}
\affiliation{Peking University, Beijing}
\affiliation{RIKEN BNL Research Center, Upton, New York 11973}
\affiliation{University of Science and Technology of China, Hefei}
\affiliation{Seoul National University, Seoul}
\affiliation{Shinshu University, Nagano}
\affiliation{Sungkyunkwan University, Suwon}
\affiliation{University of Sydney, Sydney NSW}
\affiliation{Tata Institute of Fundamental Research, Bombay}
\affiliation{Toho University, Funabashi}
\affiliation{Tohoku Gakuin University, Tagajo}
\affiliation{Tohoku University, Sendai}
\affiliation{Department of Physics, University of Tokyo, Tokyo}
\affiliation{Tokyo Institute of Technology, Tokyo}
\affiliation{Tokyo Metropolitan University, Tokyo}
\affiliation{Tokyo University of Agriculture and Technology, Tokyo}
\affiliation{Virginia Polytechnic Institute and State University, Blacksburg, Virginia 24061}
\affiliation{Yonsei University, Seoul}
  \author{K.~Abe}\affiliation{High Energy Accelerator Research Organization (KEK), Tsukuba} 
  \author{I.~Adachi}\affiliation{High Energy Accelerator Research Organization (KEK), Tsukuba} 
  \author{H.~Aihara}\affiliation{Department of Physics, University of Tokyo, Tokyo} 
  \author{D.~Anipko}\affiliation{Budker Institute of Nuclear Physics, Novosibirsk} 
 \author{K.~Arinstein}\affiliation{Budker Institute of Nuclear Physics, Novosibirsk} 
 \author{V.~Aulchenko}\affiliation{Budker Institute of Nuclear Physics, Novosibirsk} 
  \author{T.~Aushev}\affiliation{Swiss Federal Institute of Technology of Lausanne, EPFL, Lausanne}\affiliation{Institute for Theoretical and Experimental Physics, Moscow} 
  \author{S.~Bahinipati}\affiliation{University of Cincinnati, Cincinnati, Ohio 45221} 
  \author{A.~M.~Bakich}\affiliation{University of Sydney, Sydney NSW} 
  \author{V.~Balagura}\affiliation{Institute for Theoretical and Experimental Physics, Moscow} 
% ...
\collaboration{The Belle Collaboration}

\noaffiliation

\begin{abstract}
 We report the results of a study of neutral $B$ meson decays to the $D^0
 \pi^+ \pi^-$ final state, where the $D^{0}$ is fully reconstructed. 
The results are obtained from an event sample 
containing 388~million $B\bar{B}$-meson pairs collected 
in the Belle experiment at the KEKB $e^+e^-$ collider. 
 The total branching fraction of the three-body decay
${\cal B}(\bar{B}^0\to D^0\pi^+\pi^-)
=(8.4\pm0.4{\rm (stat.)}\pm0.8{\rm (syst.)})\times10^{-4}$ has been measured.
The intermediate resonant structure  of these three-body decays has
 been 
studied.
From a Dalitz plot analysis we have obtained   
the product of the branching fractions for $D^{*+}_2$ and $D^{*+}_0$ production: 
$
{\cal B}(\bar{B}^0\to D^{*+}_2\pi^-)\times {\cal B}(D_2^{*+}\to D^{0}\pi^+)=
(2.15\pm0.17{\rm (stat.)}\pm0.29{\rm (syst.)}\pm0.12{\rm (mod.)})\times10^{-4},
$ and $
{\cal B}(\bar{B}^0\to D^{*+}_0\pi^-)\times {\cal B}(D_0^{*+}\to
D^{0}\pi^+)=(0.60\pm0.13{\rm (stat.)}\pm0.15{\rm (syst.)}\pm0.22{\rm (mod.)})\times10^{-4}.
$
This is the first observation of the $\bar{B}^0\to D^{*+}_0\pi^-$ decay.
The $\bar{B^0} \to D^0\rho^0$ and $D^0f_2$ branching fractions are 
measured to be:
$
{\cal B}(\bar{B}^0\to  D^0 \rho^0)=
(3.19\pm0.20{\rm (stat.)}\pm{0.24}{\rm (syst.)}\pm0.38{\rm (mod.)})\times10^{-4},
$
and
$
{\cal B}(\bar{B}^0\to D^0 f_2 )=
(1.20\pm0.18{\rm (stat.)}\pm0.21{\rm (syst.)}\pm0.32{\rm (mod.)})\times10^{-4}.$

\end{abstract}

\pacs{13.25.Hw, 14.40Lb, 14.40.Nd}

\maketitle

\tighten

{\renewcommand{\thefootnote}{\fnsymbol{footnote}}}
\setcounter{footnote}{0}
\section{Introduction}
\begin{figure}[h]
\begin{center}
\begin{tabular}{c}
\includegraphics[height=9 cm, width=12 cm]{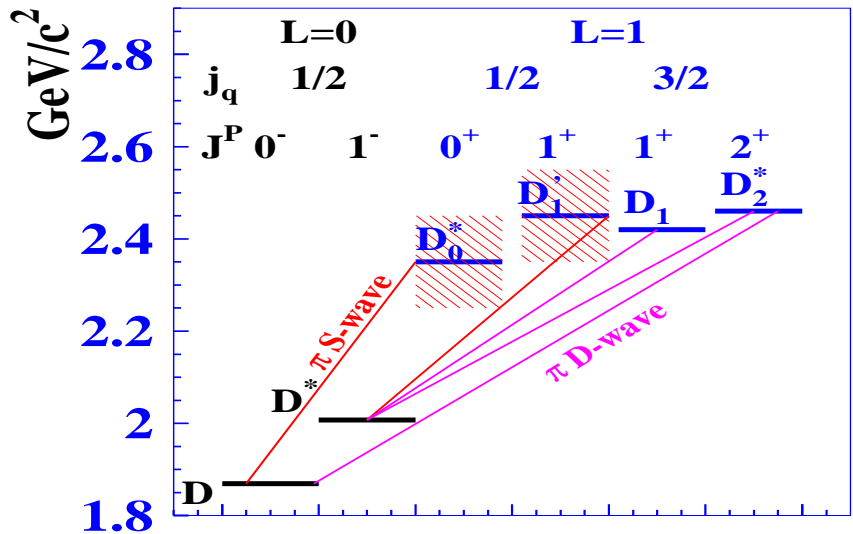}
\end{tabular}
\end{center}
\caption{Spectrum of $c\bar{q}$-meson excitations. The lines indicate possible one-pion 
transitions. The $D^*_0,~D'_1$ mesons are broad which is  indicated
by  shaded areas.}
\label{fi:spec}
\end{figure}

The decay $\bar{B}^0\to {D}^0 \pi^+ \pi^-$ includes 
intermediate states $D^{**^+}\pi^-$, where $D^{**}$'s  
are $P$-wave excitations of states 
containing one charmed and one light ($q=u,d$) quark that decay to the
$D^0\pi^+$ final state.
Figure~\ref{fi:spec} shows the spectrum and the allowed transitions of 
$c\bar{q}$-meson states. In the heavy-quark limit, the $c$-quark spin
${\vec s}_c$ decouples from the other degrees of freedom, and the total
angular momentum of the light quark ${\vec j}_q=\vec{L}+{\vec s}_q$ is a good 
quantum number.  Four $P$-wave states 
with the quantum numbers
$0^+(j_q=1/2),~1^+(j_q=1/2),~1^+(j_q=3/2)$ and $2^+(j_q=3/2)$
are expected;  these are
usually labeled as $D^*_0,~D'_1,~D_1$ and $D^*_2$, respectively.   

The two $j_q=3/2$ states have narrow widths of about 20-40~MeV 
and are well 
established~\cite{AR1,AR2,AR3,e691,CL15,e687,CL2,dobs,DELPHI,DELPHI1,ALEPH}.
 The measured masses 
agree with model predictions~\cite{isgur,rosner,godfrey,falk}.
The remaining $j_q=1/2$ states are expected to be broad and decay 
via $S$-waves.
The $B\to D^{}\pi\pi$ decay process provides a  way to  study
$D^{**}$ production.  Angular analysis of the decay products can
be used to determine  $D^{**}$ meson quantum numbers. 
These results also provide a test of Heavy Quark Effective Theory (HQET) and
QCD sum rules~\cite{QCDSR1,QCDSR}.

A study of neutral $D^{**0}$ production in charged $B$-decays has been 
recently reported by Belle~\cite{mybelle}, 
where four $D^{**}$ states are 
observed and the production rates of the broad ($j=1/2$) states 
are found to be of the same order-of-magnitude as those for the 
narrow ($j=3/2$) states.
This paper describes an analysis of the $B^0\to \bar{D}^0 \pi^+ \pi^-$ decay
that is performed in a manner similar to that of the previous Belle analysis
of the
$\bar{B}^0\to D^{**+}\pi^-$ decay~\cite{hhh}. The results presented
here supersede those of Ref.~\cite{hhh}. 

\begin{figure}[h]
\begin{center}
\begin{tabular}{ccc}
\includegraphics[width=5 cm]{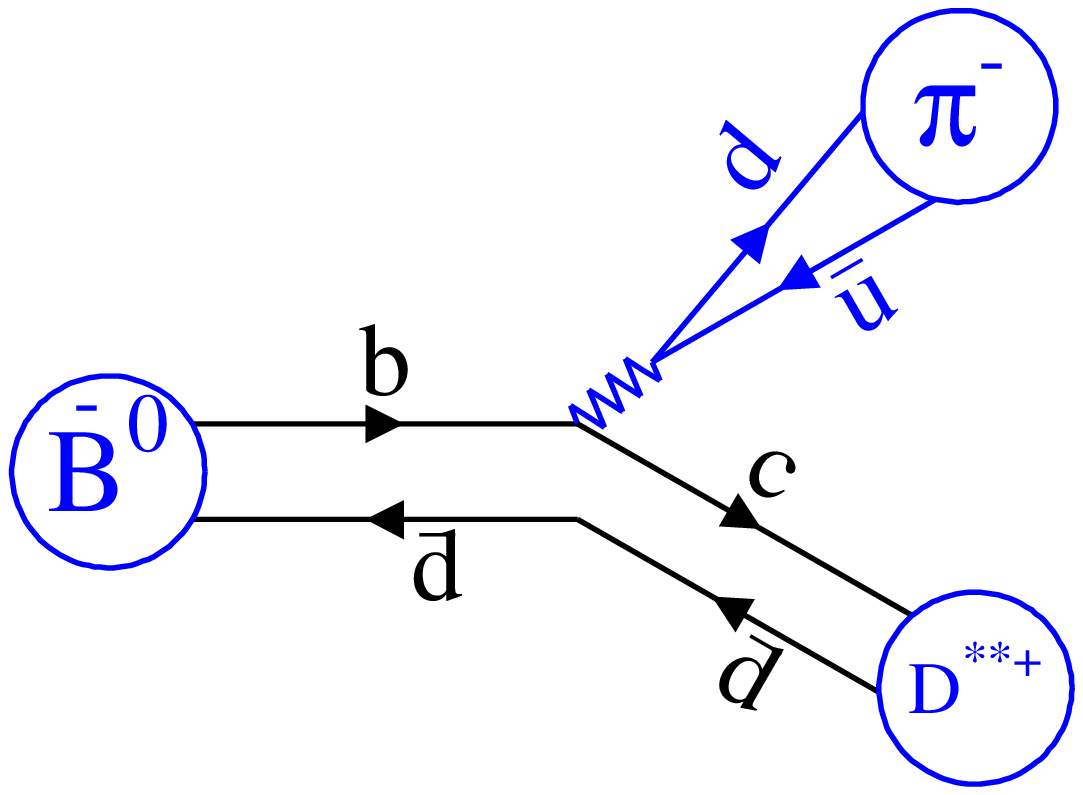}&
\includegraphics[width=5 cm]{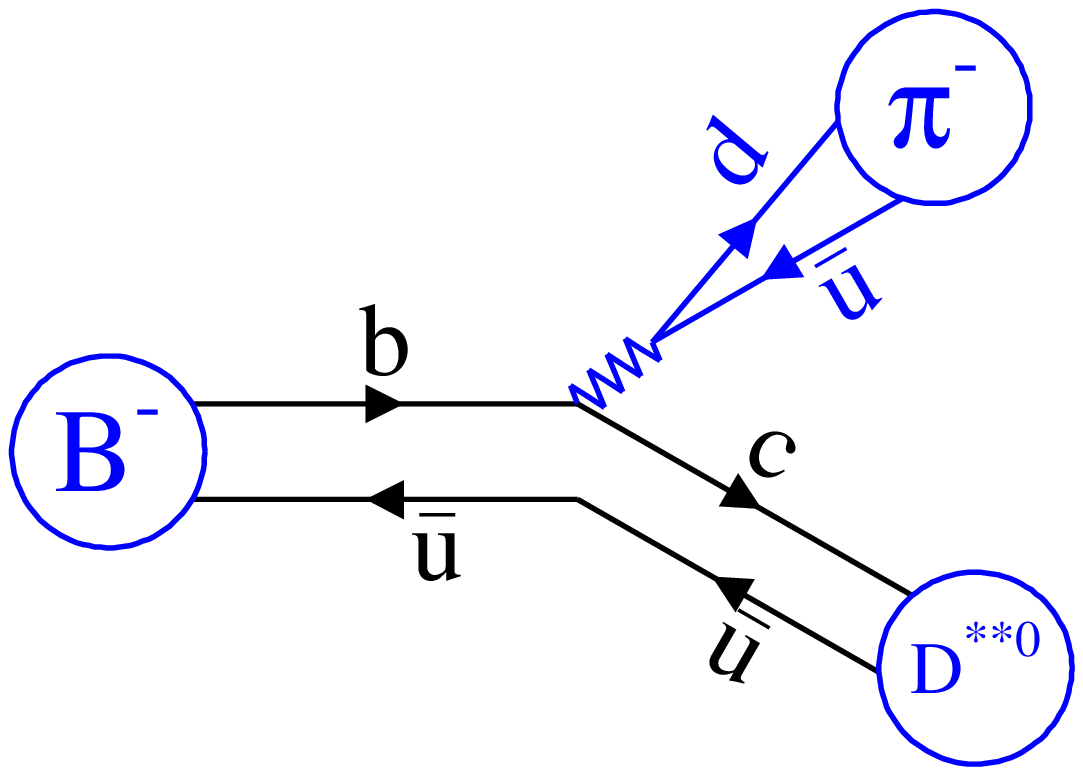}&
\includegraphics[width=5 cm]{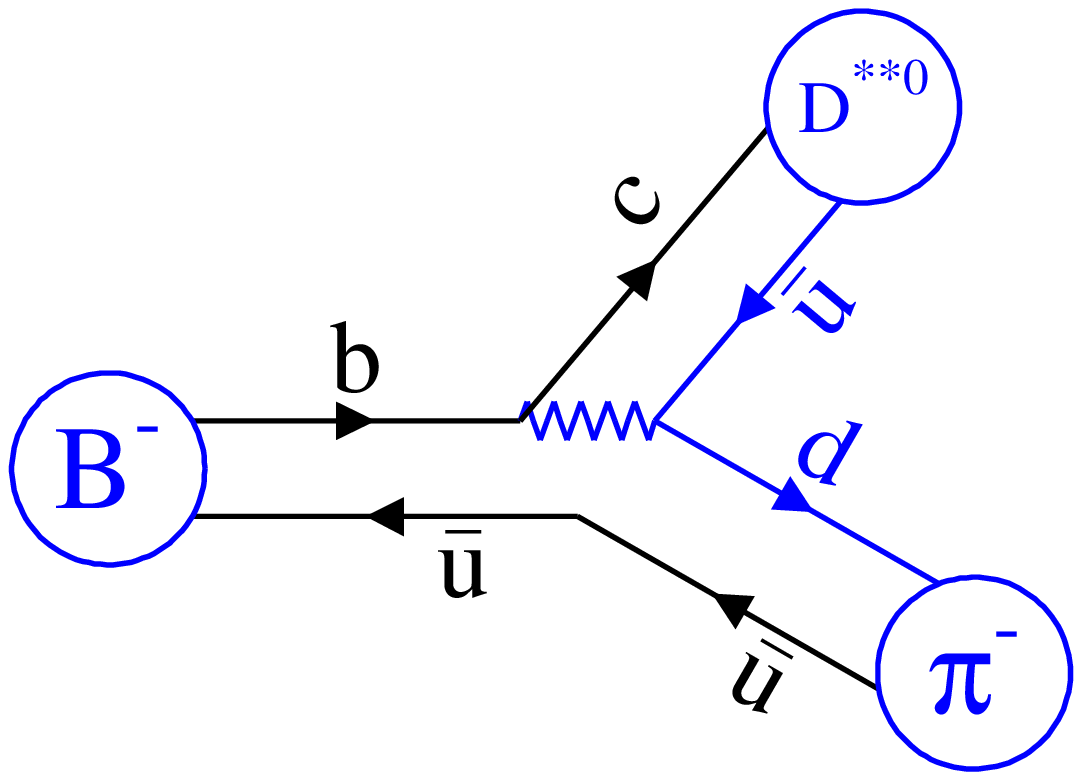}\\
\vspace*{-4 cm} & \\
{\bf\large \hspace*{-1cm} a)}&{\bf\large \hspace*{-1cm} b)}&{\bf\large
  \hspace*{-1cm} c)}\\ 
\vspace*{3 cm} & \\
\end{tabular}
\caption{Quark-line diagrams for  neutral (a) and charged (b and c) $B$ decays.} 
\label{f:fd1}
\end{center}
\end{figure}

The neutral  $B$ decay to $D^{**}\pi$ is described by 
 the tree diagram only as shown in Fig.~\ref{f:fd1}(a) while 
for  the charged $B$ decay to $D^{**}\pi$, the amplitude
receives contributions from both tree and 
color-suppressed diagrams as shown in Fig.~\ref{f:fd1}(b,c).
$D^{**}$ tree-diagram production amplitudes are described
by the Isgur-Wise functions $\tau_{1/2}$ and  $\tau_{3/2}$.
According to the QCD sum rule~\cite{QCDSR1,QCDSR}, $\tau_{1/2}\ll \tau_{3/2}$ and
one would expect suppression of decays to the broad state.
The observation that  
the production rates of the broad ($j=1/2$)-states are comparable with
those of the narrow  ($j=3/2$)-states indicates either a large
contribution of the color-suppressed diagram or the violation of the
sum rule above. Measurement of the decay rates of the neutral $B$
allows one to 
test the contribution of the tree-diagram only and also test
the QCD sum rule.

In this analysis, the final state contains 
two pions of opposite sign, and these can originate from resonant
states such as the $\rho^0,~f_0,~f_2$, etc.  While the possible
presence of $\pi\pi$ resonant structures complicates the analysis, it
can also provide valuable information about the 
mechanism of these decays.

\section{The Belle detector}
  The Belle detector~\cite{Belle} is a large-solid-angle magnetic spectrometer
that consists
of a silicon vertex detector (SVD), 
a 50-layer central drift chamber
(CDC) for charged particle tracking and specific ionization measurement 
($dE/dx$), an array of aerogel threshold \v{C}erenkov counters (ACC), 
time-of-flight scintillation counters (TOF), and an array of 8736 CsI(Tl) 
crystals for electromagnetic calorimetry (ECL) located inside a superconducting
solenoid coil that provides a 1.5~T magnetic field. An iron flux return located
outside the coil is instrumented to detect $K_L^0$ mesons and identify muons
(KLM). 
We use a GEANT-based Monte Carlo (MC) simulation to
model the response of the detector and determine its acceptance~\cite{sim}.

  Separation of kaons and pions is accomplished by combining the responses of 
the ACC and the TOF with $dE/dx$ measurements in the CDC
 to form a likelihood $\cal{L}$($h$) where $h=(\pi)$ or $(K)$. 
Charged particles 
are identified as pions or kaons using the likelihood ratio 
(${\mathcal R}$):
\[{\rm {\mathcal R}}(K)=\frac{{\cal{L}}(K)}{{\cal{L}}(K)+{\cal{L}}(\pi)};~~
{\rm {\mathcal R}}(\pi)=
\frac{{\cal{L}}(\pi)}{{\cal{L}}(K)+{\cal{L}}(\pi)}=1-{\rm {\mathcal R}}(K).\]
 A more detailed
description of the Belle particle identification can be found in 
Ref.~\cite{PID}.

\section{Event selection}

A data sample of 357~fb$^{-1}$ 
(388 million~$B\bar{B}$ events) collected at the 
$\Upsilon (4S)$ resonance  is used in this analysis.
Candidate $\bar{B}^0\to D^0\pi^+\pi^-$ 
events are selected, where
the $D^0$  mesons decay via the $D^0\to K^-\pi^+$ mode.
The signal-to-noise ratios for other $D^0$ decay modes are found to be
significantly lower and, therefore, 
are not used.
(The inclusion of charge conjugate states is implied by default
throughout this paper.) 

Charged tracks are selected with requirements based on the  
average hit residuals and impact parameters relative to the interaction
point. We require that the polar angle of each track be in
the angular  range of $17^{\circ}-150^{\circ}$ and that the 
track transverse momentum be
 greater than 50 MeV/$c$ for kaons and greater than 25 MeV/$c$ for pions.

Charged kaon candidates are identified by the requirement 
${\rm {\mathcal R}}(K)>0.6$, 
which has an efficiency of $90\%$ and a pion misidentification probability of
approximately  $10\%$.
For pion candidates we require \mbox{${\rm {\mathcal R}}(\pi)>0.2$}.
Kaon and pion candidates are rejected if the track is positively 
identified as an electron.

Candidate $D^{0}$ mesons are
$K^{-}\pi^{+}$ combinations with an invariant mass 
within $\pm$12~MeV/c$^2$ of the nominal 
$D^0$ mass, which corresponds to $\sim$2.5\,$\sigma_{K\pi}$.
We reject $D^0$ candidates that, when combined with any $\pi^0$
in the event, has a value of $M_{D\pi^0}-M_{D^0}$ that is within
$\pm2.5~{\rm MeV}/c^2$ of the nominal $D^{*0}$-$D^0$ mass difference.

$B$ meson candidates are identified 
by their center-of-mass (c.m.)\ energy difference 
$\Delta E=(\sum_iE_i)-E_{\rm b}$, and the
beam-constrained mass 
$M_{\rm bc}=\sqrt{E^2_{\rm b}-(\sum_i\vec{p}_i)^2}$, where 
$E_{\rm b}=\sqrt{s}/2$ is the beam energy in the $\Upsilon(4S)$
c.m.\ frame, and $\vec{p}_i$ and $E_i$ are the c.m.\ three-momenta and 
energies, respectively of the $B$ meson candidate decay products. We
select events satisfying
$M_{\rm bc}>5.25$~GeV/$c^2$ and $|\Delta E|<0.10$~GeV.

To suppress the large continuum background ($e^+e^-\to q\bar{q}$,
where $q=u,d,s,c$), topological variables are used. Since  
the produced $B$ mesons 
are almost at rest in the c.m. frame, the signal
event shapes tend to be 
isotropic while  continuum $q\bar{q}$ events
tend to have  a two-jet structure. We use the angle between the thrust axis of 
the $B$ candidate and that of the rest of the event ($\Theta_{\rm thrust}$)
to discriminate between these two cases. The distribution of
$|\cos\Theta_{\rm thrust}|$ is strongly peaked near $|\cos\Theta_{\rm thrust}|=1$
for $q\bar{q}$ events and is nearly flat for  $\Upsilon(4S)\to
B\bar{B}$ events.
We require $|\cos\Theta_{\rm thrust}|<0.8$, which eliminates about 
83$\%$ of the continuum background while retaining about 80$\%$ of signal 
events.

There are events for which two or more track combinations pass all
the selection criteria. According to MC simulation,
this occurs  primarily
because of the misreconstruction of the low momentum pion from
$D^{**}\to D^{}\pi$ decays. To avoid multiple entries, 
the combination that has the  minimum difference of 
$z$ coordinates at the interaction point, $|z_{\pi_1}-z_{\pi_2}|$, 
of the tracks corresponding to the pions from
$B\to D\pi_1\pi_2$  are
selected~\cite{foot1}.
 This selection also
suppresses  combinations that
include pions from $K_S^0$ decays. In the case of multiple $D\to K\pi$ 
combinations, the one with the invariant
mass closest to the $D^0$ mass is selected.  
 
\section{${\boldmath \bar{B}^{0}\to D^{0}\pi^{+}\pi^{-}}$ branching fraction}
The $D^{0}\pi^{+}\pi^{-}$ final state,
together with three-body  and quasi-two-body contributions,
includes  the two-body $\bar{B}^{0}\to D^{*+}\pi^{-}$ decay followed by the 
decay $D^{*+}\to D^{0}\pi^{+}$. 
We obtain the branching fraction of the three-body decay   $\bar{B}^{0}\to
D^{0}\pi^{+}\pi^{-}$ excluding the contribution of  $\bar{B}^{0}\to D^{*+}\pi^{-}$.
Using the $M_{D\pi}-M_{D}$ mass difference, 
we subdivide the total sample in to two subsamples as follows. Events that have 
a $D\pi$ combination with $M_{D\pi}-M_{D}$ within 
$3~{\rm MeV/c}^2$  ($\sim6\sigma$) of the nominal
$D^{*+}-D^0$ mass difference
are denoted  below as sample (2);  the rest 
of the $D\pi\pi$ events are denoted as sample (1).  Sample (2) is used to 
crosscheck our procedures.

\begin{figure}[h]
\begin{tabular}{cc}
 \includegraphics[height=8 cm]{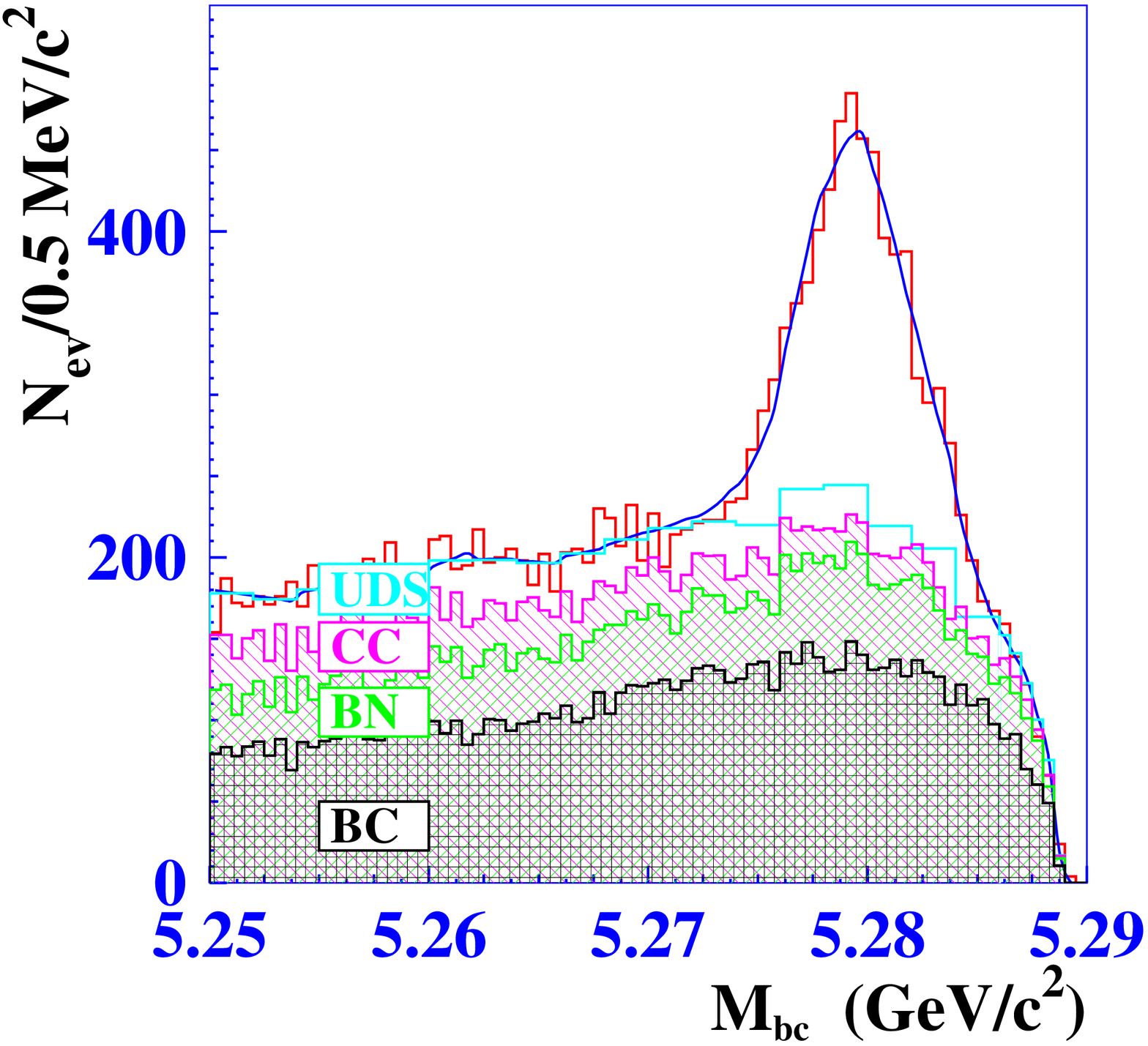}&
 \includegraphics[height=8 cm]{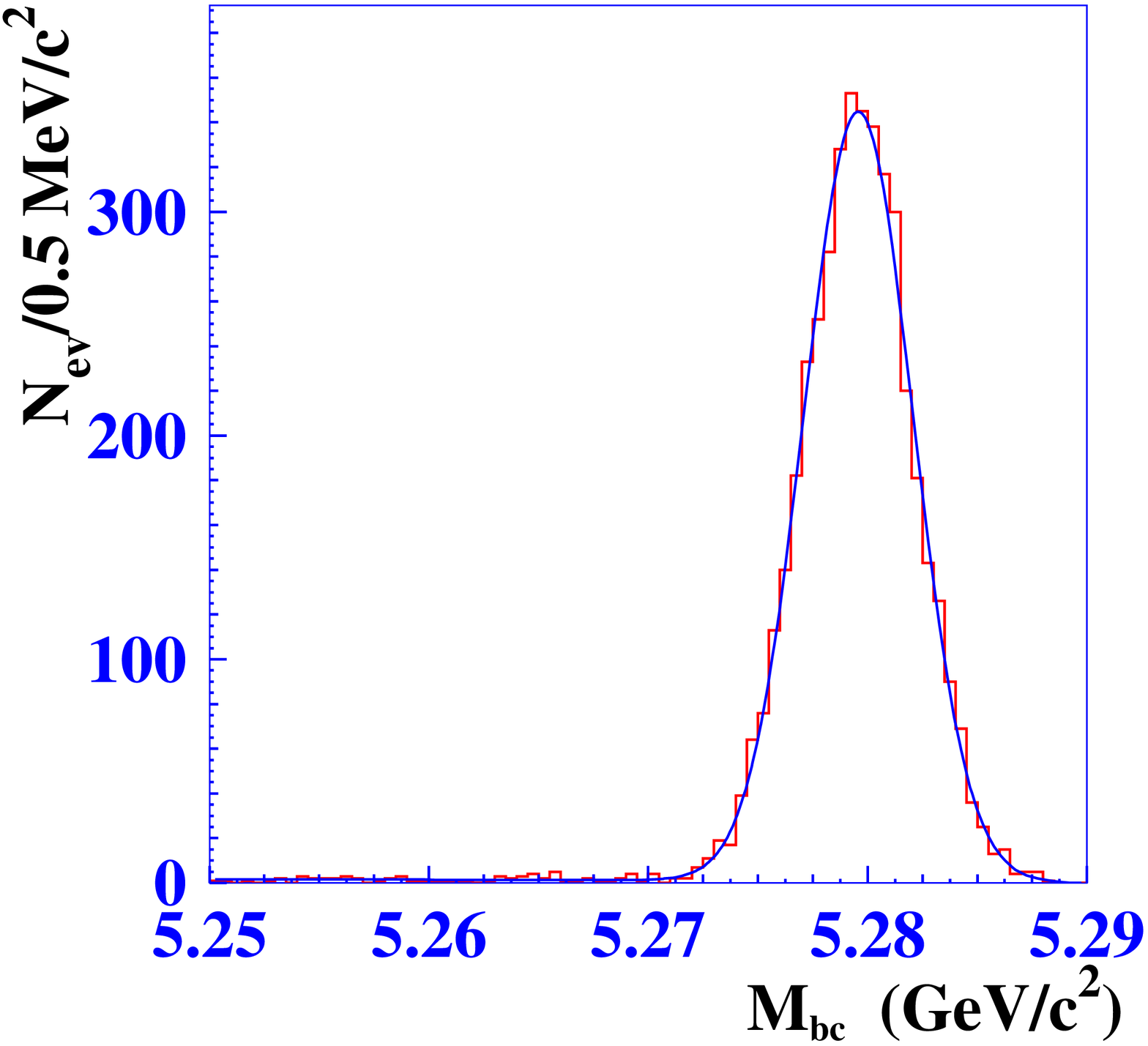}\\
 \includegraphics[height=8 cm]{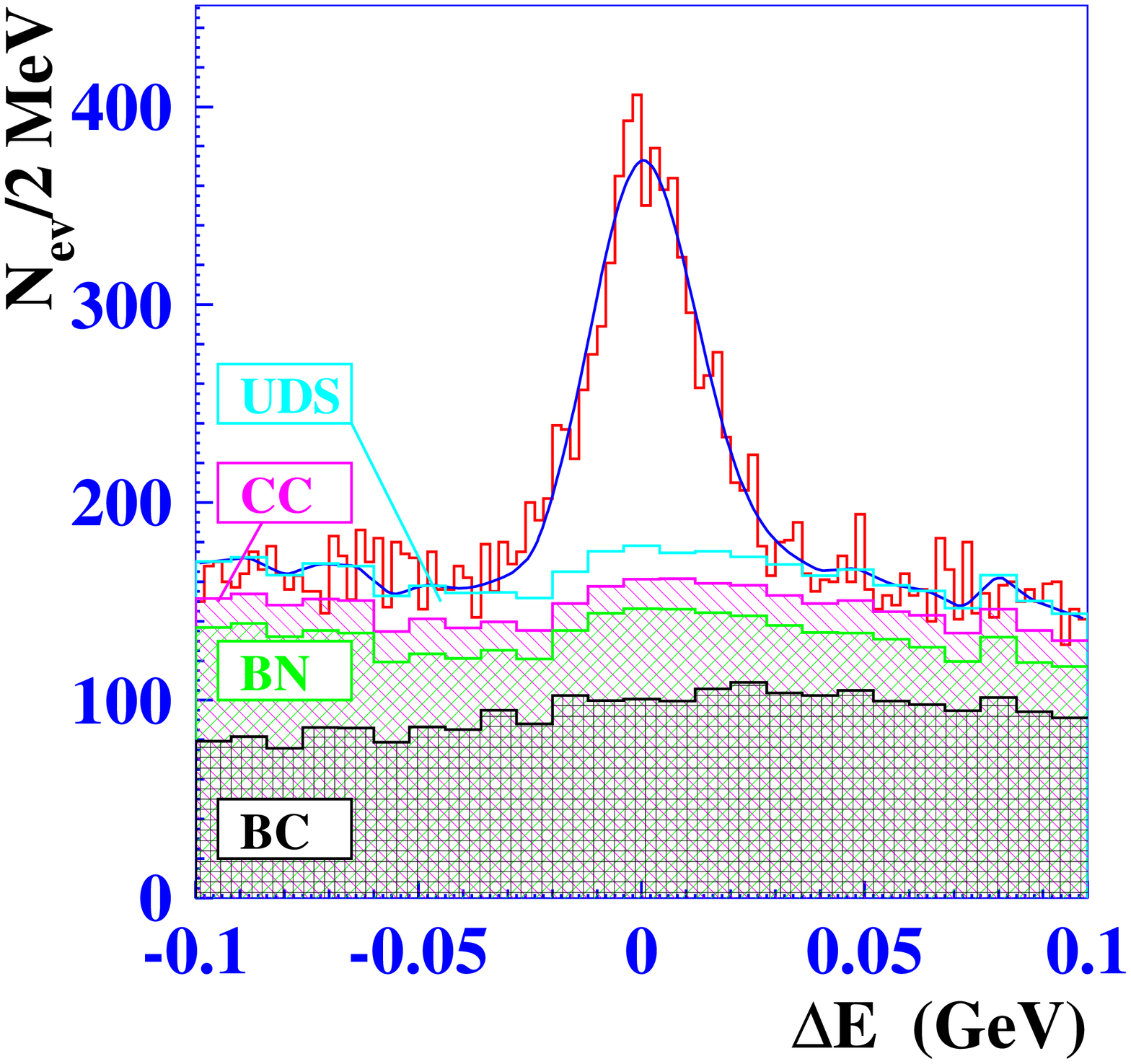}&
 \includegraphics[height=8 cm]{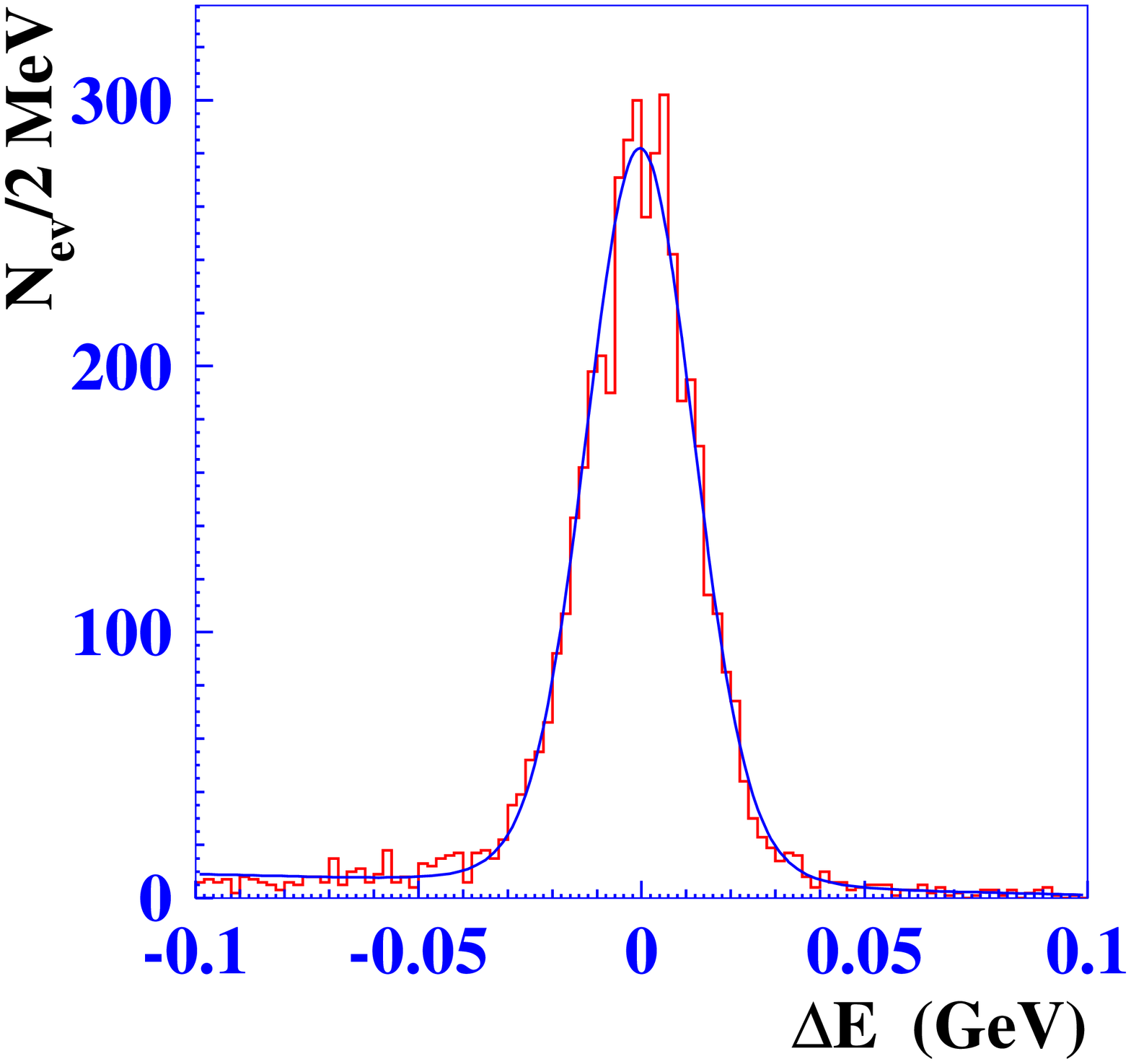}\\
\vspace*{-16 cm} & \\
{\bf\large \hspace*{-3cm} a)}&{\bf\large \hspace*{-3cm} b)}\\
\vspace*{7. cm} & \\
{\bf\large \hspace*{-3cm} c)}&{\bf\large \hspace*{-3cm} d)}\\
\vspace*{6.5 cm} & \\
\end{tabular}
\caption{$M_{\rm bc}$ and  $\Delta E$ distributions for
 $\bar{B}^{0}\to D^{0}\pi^{+}\pi^{-}$ events.
Sample~(1) distributions are shown in (a) and (c); those for sample~(2)
are shown in (b) and (d).
 }
\label{f:dpmbde}
\end{figure}

 The $M_{\rm bc}$  and $\Delta E$ distributions
for $\bar{B}^{0}\to D^{0}\pi^{+}\pi^{-}$ events 
are shown in Fig.~\ref{f:dpmbde}. 
The distributions are plotted for events that satisfy 
the selection criteria for the other variable: 
$|\Delta E|<25$~MeV and $|M_{\rm bc}-m_B|<7$~MeV/$c^2$ for the 
$M_{\rm bc}$ and  $\Delta E$ histograms, respectively, where $m_B$
is the nominal $B$ mass. Distinct signals 
are evident in all of the distributions.

The background shape is obtained from generic MC data samples that include
$B^+B^-$ (BC) and $B^0\bar{B^0}$ (BN),
continuum charm production (CC) and continuum with light
quarks (UDS), each corresponding to approximately twice the luminosity of the 
experimental data.
The $D\pi$ and $\pi\pi$ invariant mass distributions are different for 
different MC samples. 
The branching fractions
used in the generic MC are measured with some experimental uncertainty
and may not reproduce the experimental data.
To improve the quality of the MC spectra, relative weights 
of these four
 components
are determined from a fit to
a two-dimensional Dalitz plot distribution for events in the $\Delta E$
sideband shown in Fig.~\ref{f:mbc}.
 The fitting function represents the sum of the 
four two-dimensional histograms with floating weights.
Each histogram is determined  from its respective MC sample. 
The weights obtained for the four
components are:
$a_{BC}=1.10\pm0.07,~a_{BN}=1.37\pm0.22,~a_{CC}=0.52\pm0.12,~a_{UDS}=0.92\pm0.22$.
The $\Delta E$ background shape is described as
$F_{bg}(\Delta E)=\sum_{i}a_iF_i(\Delta E)$, where
$F_i(\Delta E)$ 
is the $\Delta E$ distribution of the i-th component obtained
from the MC sample.

The signal yield is obtained  by fitting the  $\Delta E$ 
distribution to the  sum of  two Gaussians
with the same mean value to describe the signal, plus the 
above-described background 
function $F_{bg}(\Delta E)$.
The width of the broader Gaussian and the relative normalization of 
the two Gaussians are fixed to the values obtained from a MC simulation;  
the signal and background normalization as well as the width of the narrow Gaussian
are left as free parameters.

The fitted signal yields  are $2909\pm115$ events and $4202\pm67$ events
for samples (1) and (2), respectively.
The reconstruction efficiencies $(23.4\pm 0.4)\%$ and $(19.0\pm 0.4)\%$
are determined from a MC
simulation that uses a Dalitz plot 
distribution that is 
 generated according to the model described 
in the next section.
Taking into account  
${\cal B}(D^0\to K^-\pi^+)=(3.80\pm0.07)\%$~\cite{PDG},   
we obtain the following branching fraction: 
$$
{\cal B}(\bar{B}^0\to D^0\pi^+\pi^-) =(8.4\pm0.4\pm0.8)\times10^{-4},
$$
where the first error is statistical and second error is systematic. 
Various contributions to the systematic error are         
listed in Table~\ref{t:sys} for both samples.  They 
include tracking efficiency,
 particle identification efficiency,
 limited MC statistics, and background uncertainty.
 The background uncertainty is obtained by varying the relative
 weights within their errors. The contribution of the non-resonant 
$\bar{B^0} \to K^-\pi^+\pi^-\pi^+$ is estimated using the $K\pi$ mass
 sidebands of the $D$ mass region and is negligible.

\begin{table}
\begin{center}
\begin{tabular}{lll}
\hline
                &sample(1)           & sample(2)\\
\hline		 	 	      	  		       
Particle identification                               & $5\%$   & $5\%$   \\        
Background uncertainty                        & $5\%$   & $1\%$   \\  
Tracking efficiency                         & $4.4\%$ & $5.4\%$ \\
MC statistics                    	           & $3\%$  & $3\%$    \\  
${\cal B}(D,D^*)$ uncertainty                 &$2.4\%$  &$2.5\%$  \\  
\hline		      	               	    	          
Total                 	           & $9.2\%$& $8.1\%$  \\
\hline
\end{tabular}
\end{center}
\caption{The systematic uncertainties for the $\bar{B}^0\to D^0\pi^+\pi^-$
branching fraction measurement.}
\label{t:sys}
\end{table}

The value of ${\cal B}(\bar{B}^0\to D^0\pi^+\pi^-)$ improves and
supersedes previous Belle result
$
{\cal B}(\bar{B}^0\to D^0\pi^+\pi^-)=(8.0\pm1.6)\times10^{-4}
$~\cite{Asish}.
The value of the branching fraction ${\cal B}(\bar{B}^0\to
  D^{*+}\pi^-)$ is obtained using  sample (2) and the PDG value ${\cal
    B}(D^{*+}\to D^0\pi^+)=(67.7\pm0.5)\%$~\cite{PDG}. The result is
$
{\cal B}(\bar{B}^0\to D^{*+}\pi^-)=(2.22\pm0.04\pm0.19)\times10^{-3},
$
which is somewhere lower than the
CLEO result
 ${\cal B}(\bar{B}^0\to D^{*+}\pi^-)=(2.81\pm0.25)\times10^{-3}$~\cite{CLEOd1}.

\subsection{$B\to D\pi\pi$ Dalitz plot analysis}
For a three-body decay of a spin zero particle, two 
variables are required to describe the decay kinematics; 
we use the 
$D^0\pi^+$ and $\pi^+\pi^-$ invariant masses squared, $M^2_{D\pi}$ and $ M^2_{\pi\pi}$,
respectively.

To analyze  the dynamics of $B\to D\pi\pi$ decays, sample~(1)  
events 
with $\Delta E$ and $M_{\rm bc}$
within the signal region
$((\Delta E+\kappa(M_{\rm bc}-m_B))/\sigma_{\Delta E})^2+((M_{\rm
  bc}-m_B)/\sigma_{M_{\rm bc}})^2<4$ 
are selected. The parameters $\sigma_{\Delta E}=11~{\rm
  MeV},~\sigma_{M_{\rm bc}}=2.7~{\rm MeV}/c^2$ and $\kappa=0.9$  are determined
from a fit to experimental data; the coefficient $\kappa$ accounts
for the correlation between  $M_{\rm bc}$ and $\Delta E$.  

\begin{figure}[h]
\begin{center}
\includegraphics[width=8 cm]{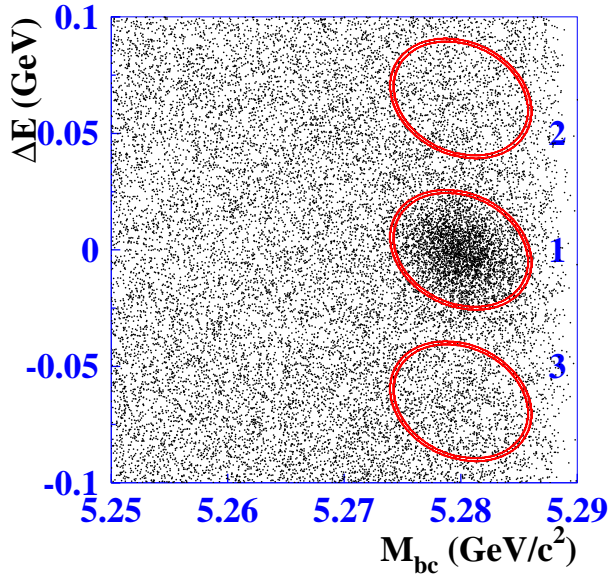}
\caption{The experimental distribution in the $ (M_{\rm bc}$-$\Delta
  E)$ plane. The ellipses show the signal (1) and sideband regions (2 \& 3).}
\label{f:mbc}
\end{center}
\end{figure}

To test and correct the shape of the background, we use
events from the $\Delta E$ sidebands, which are defined as:
$((\Delta E\pm65\,{\rm MeV}+\kappa(M_{\rm bc}-m_B))/\sigma_{\Delta E})^2+((M_{\rm
  bc}-m_B)/\sigma_{M_{\rm bc}})^2<4$. Figure~\ref{f:mbc} shows the 
signal and sideband regions in the $ M_{\rm bc}$-$\Delta E$ plane.

\begin{figure}[h]
\begin{center}
\begin{tabular}{cc}
\includegraphics[width=8 cm]{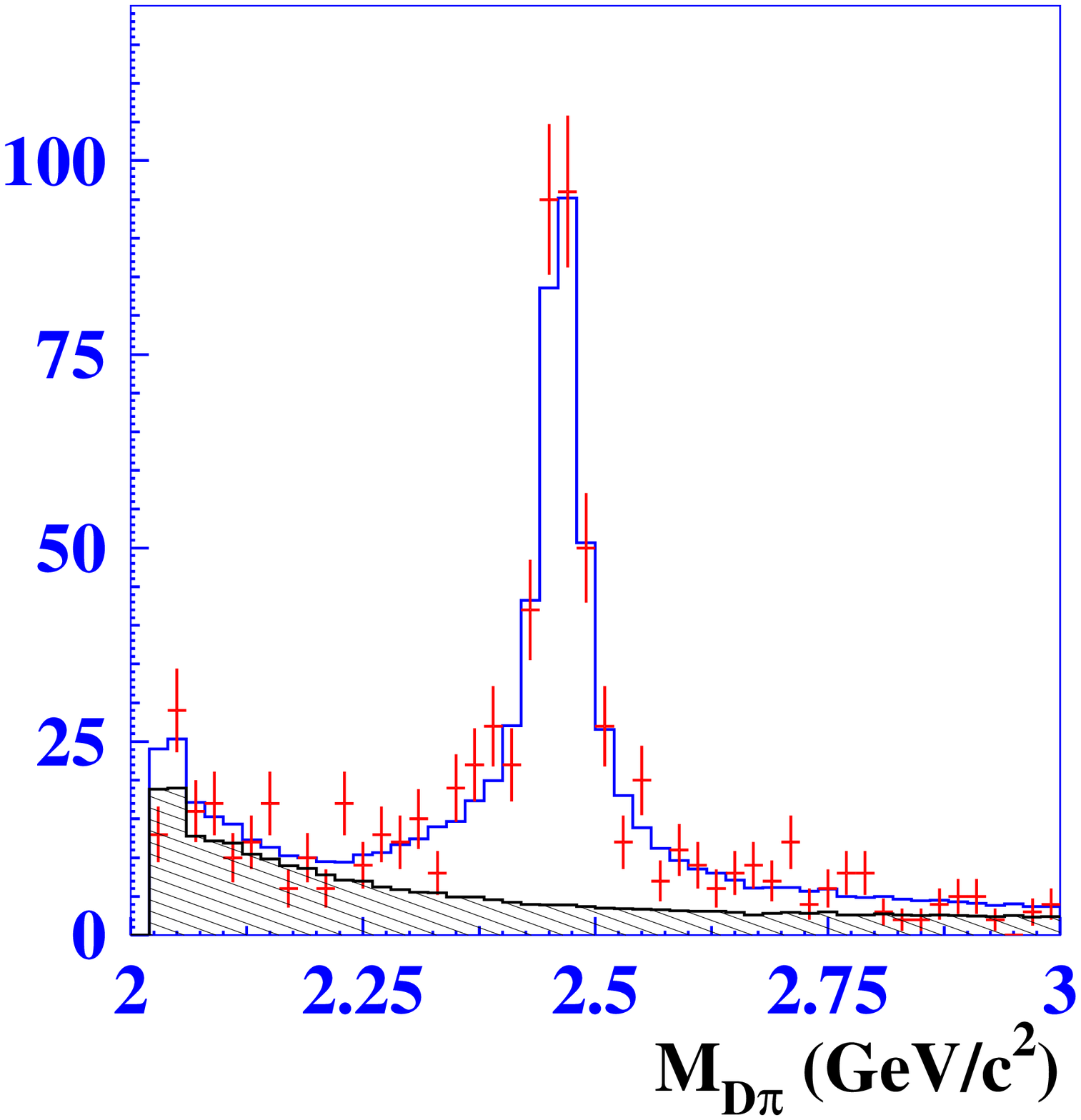}&
\includegraphics[width=8 cm]{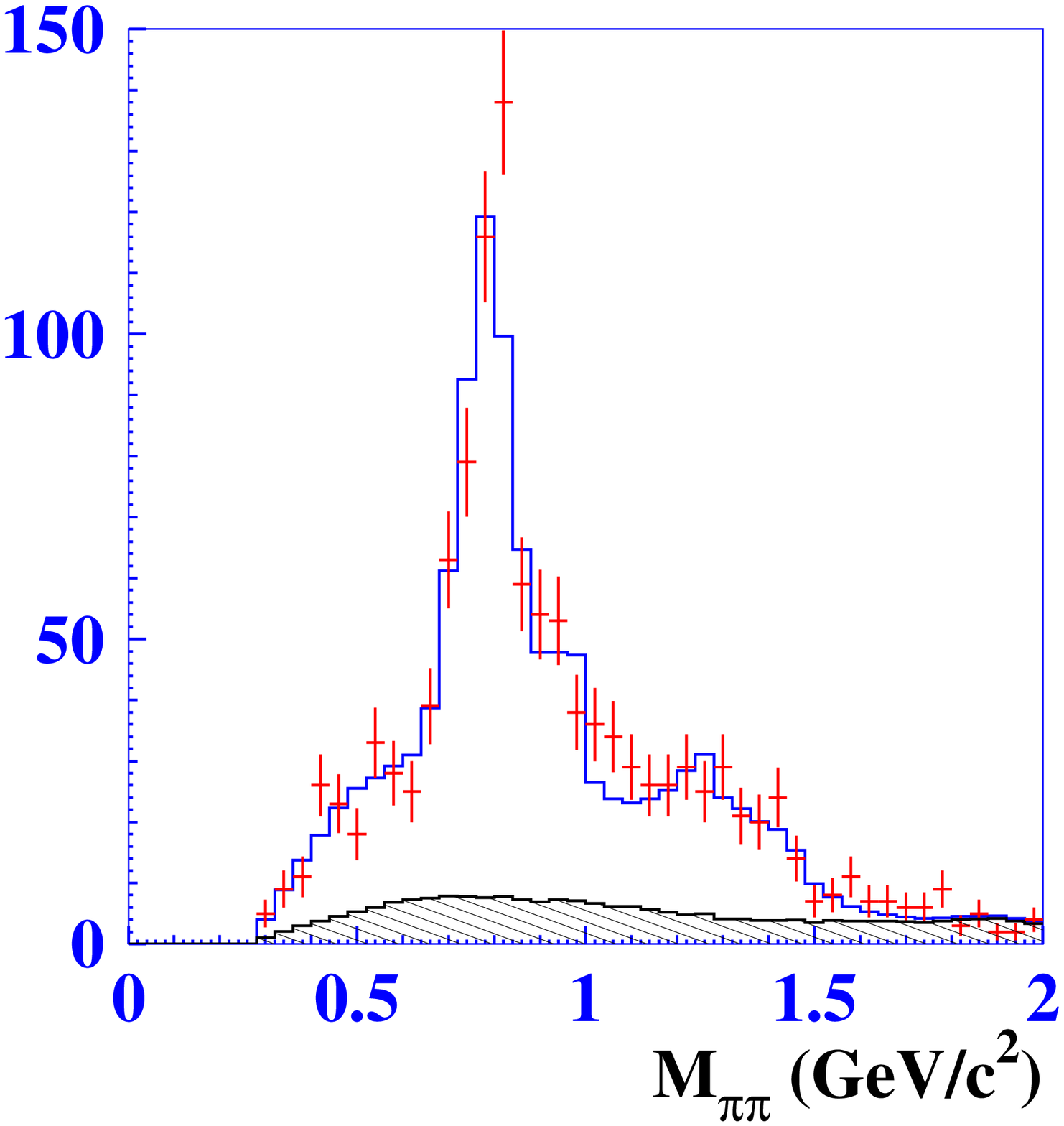}\\
\vspace*{-8 cm} & \\
{\bf\large \hspace*{-3cm} a)}&{\bf\large \hspace*{-3cm} b)}\\
\vspace*{6.5 cm} & \\
\end{tabular}
\caption{$D\pi$ (a) and $\pi\pi$ (b) mass distributions for
  sample (1) events with helicity angles  $\cos\theta_{h}>0$.
 The points
  with error bars
  correspond to the signal region events,  the hatched 
  region indicates the background obtained from generic MC events normalized
  to the sideband data. The open histogram shows the
  fitted function after efficiency correction.
}
\label{f:dpp_M}
\end{center}
\end{figure}

The $D\pi$ and $\pi\pi$ mass distributions for the signal and sideband
events  (sample 1)
are shown in Fig.~\ref{f:dpp_M}. In the  $D\pi$ mass distribution
a narrow peak corresponding to $D_2^*$ is evident.  The $\pi\pi$ mass distribution
shows a peak corresponding to the $\rho$ meson as well as a 
structure at the $1.2-1.3\,\rm GeV/c^2$ mass region that is 
presumably  due to $f_0(1370)$ and
$f_2(1270)$ production.

\begin{figure}[h]
\begin{center}
\begin{tabular}{cc}
\includegraphics[width=8 cm]{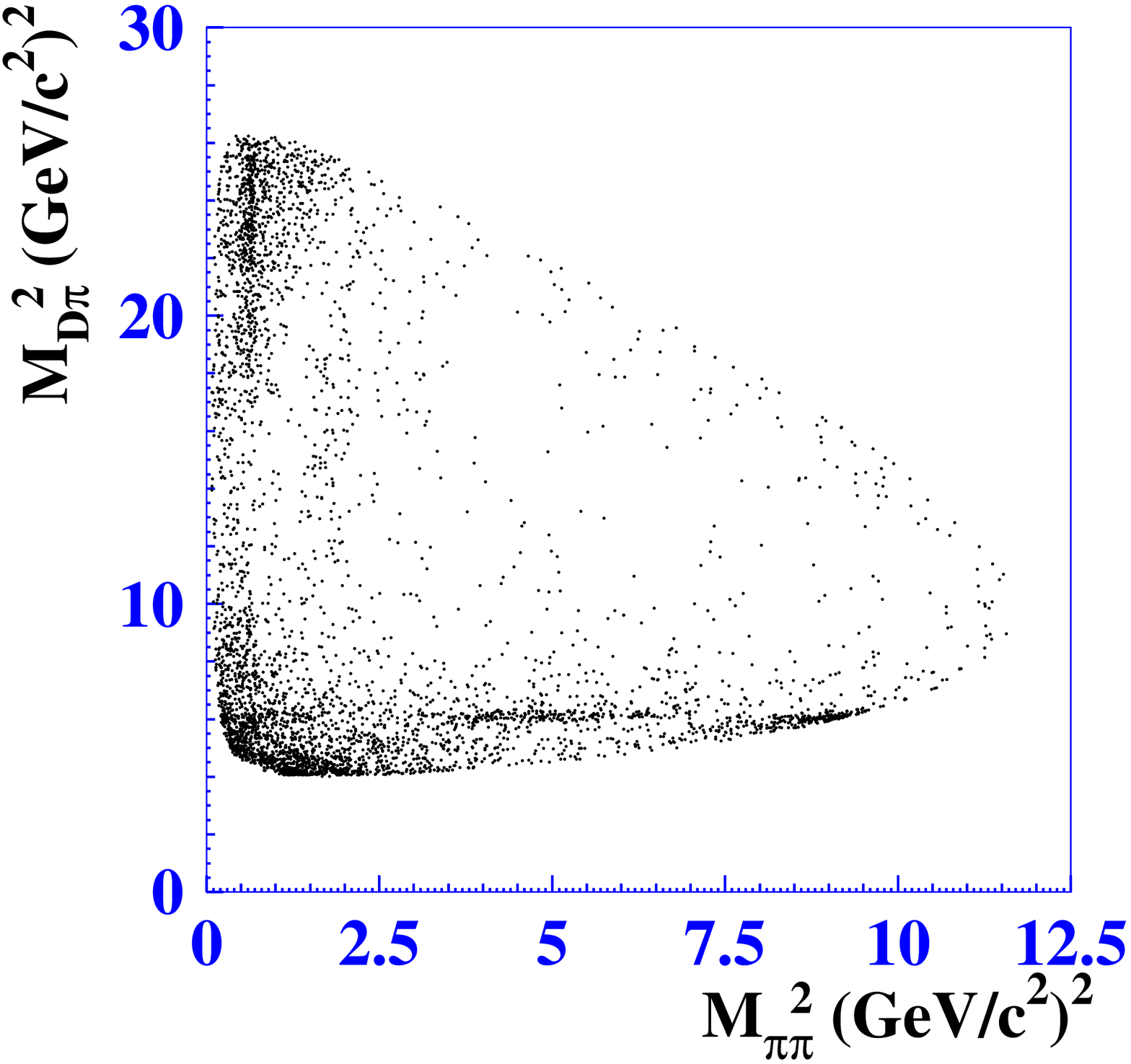}&
\includegraphics[width=8 cm]{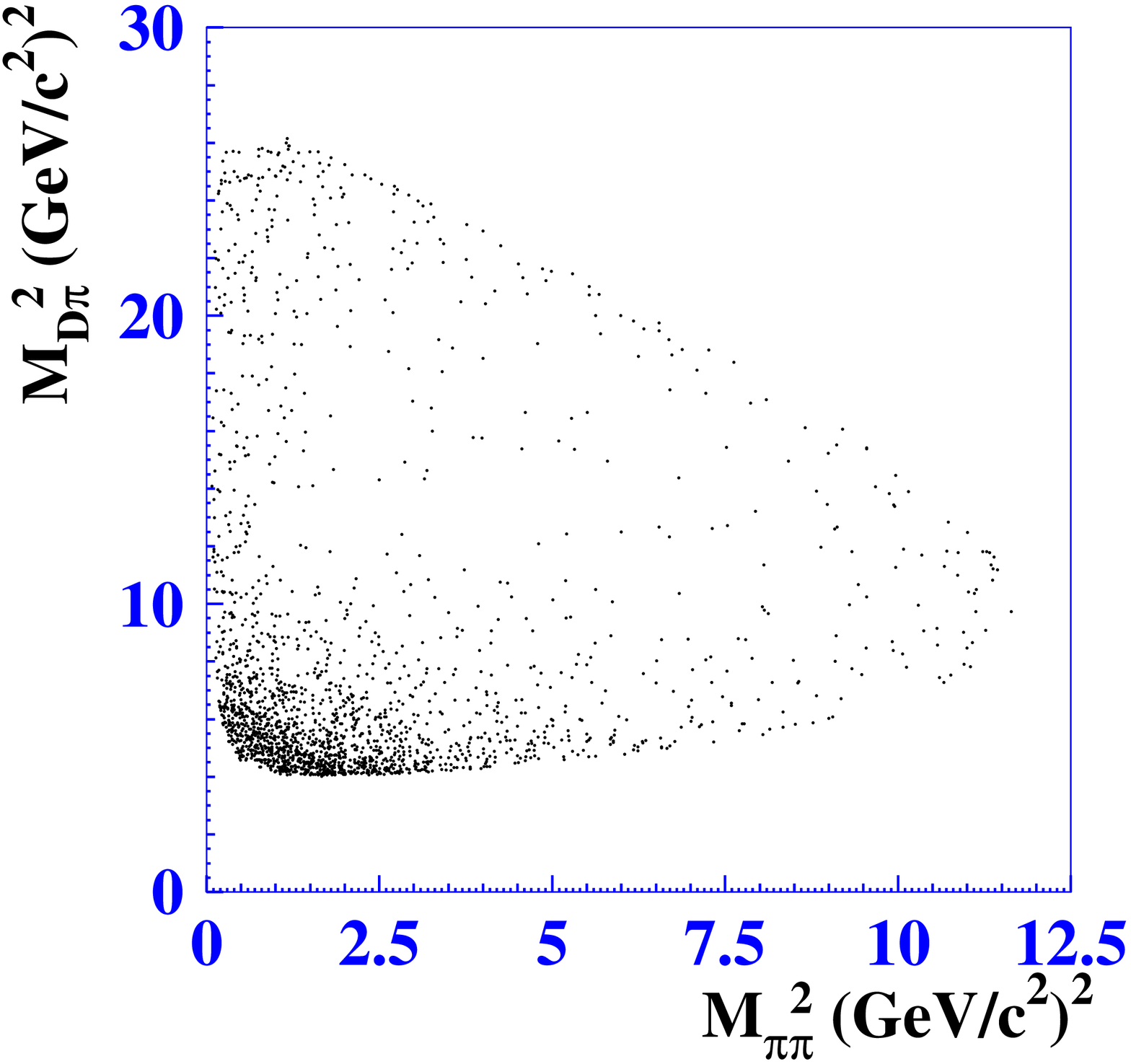}\\
\vspace*{-8 cm} & \\
{\bf\large \hspace*{-3cm} a)}&{\bf\large \hspace*{-3cm} b)}\\
\vspace*{6.5 cm} & \\
\end{tabular}
\caption{The Dalitz plot for  (a) signal events; (b) sideband events.}
\label{f:dpp_DP}
\end{center}
\end{figure}

The 
$M^2_{D\pi}$ and $M^2_{\pi\pi}$
 Dalitz plot distributions
for the signal and sideband regions are shown 
in Fig.~\ref{f:dpp_DP}.
The Dalitz plot boundary is fixed by the
decay kinematics and the masses of the daughter particles. 
In order to have the same Dalitz plot boundary for  
both signal and sideband event samples, fits 
where the  $K\pi$ mass is constrained to $M_D$ and $D\pi\pi$ mass to 
$m_B$ are performed. 
The mass-constrained fits also slightly improve the
accuracy of 
$M^2_{D\pi}$ and $M^2_{\pi\pi}$.

To extract the amplitudes and phases of different intermediate states,
an   unbinned fit to the Dalitz plot is performed using the method
described in Ref.~\cite{mybelle}.
The event density function in the Dalitz plot includes both the signal  
and  background functions.

The backgrounds in the Dalitz plot are mostly
combinatorial and have neither resonant structure (Fig.~\ref{f:dpp_DP}b)
nor specific helicity behavior.
The background shape 
is obtained from an unbinned fit 
to  $\Delta E$ sideband events using the
 weights described above. The background Dalitz plot density is modeled by a smooth
 two-dimensional function.
The number of background events in the signal region 
is scaled according to the relative areas of the signal and  
sideband regions. 

There is no general method to describe a three-body
amplitude. In this paper we represent the $D\pi\pi$ amplitude as
the sum of Breit-Wigner function contributions for different intermediate
two-body states.  Such
an approach is not exact because it is neither analytic nor unitary and
does not take into account a complete description of the 
final state interactions.  Nevertheless, 
the sum of Breit-Wigner functions describes the main features of the
amplitude behavior and allows one to find and distinguish the
contributions of two-body intermediate states, their interference,
and the effective parameters of these states. We followed the same approach 
in the analysis of charged $B$ decays~\cite{mybelle}.

In the $D^0\pi^+\pi^-$ final state,
a  combination of the  $D^0$-meson and a pion can form a vector meson $D^{*+}$, a tensor meson 
$D^{*+}_2$ or a scalar state $D_0^{*+}$; the axial vector mesons $D_1^+$ and 
$D'^{+}_1$ cannot decay to two pseudoscalars because of angular 
momentum and parity conservation. The region of $D^{0}\pi^+$
invariant mass that
corresponds to the $D^{*+}$
is excluded from the fit by requiring 
$|M_{D\pi}-M_{D^*}|> 3$~MeV/c$^2$.   However,  in $B$-meson decay, a
virtual $D^{*+}$ (referred to as $D^*_v$)  can be produced off-shell
with $M_{D\pi}$ above
the $D^0\pi^+$ production threshold and such a process can
contribute to the amplitude. 
Another virtual hadron that can be produced in this combination is 
$B^{*-}$ (referred to as $B^{*}_v$):
$B\to B^*_v\pi$ and $B^*_v\to D\pi$. 
For the
mass of ${B}^{*-}$ as well as the mass and width of
the $D^{*+}$, we use the PDG
values~\cite{PDG}; the widths of ${B}^{*-}$ are
calculated from the width of the $D^{*-}$ in the HQET approach.
To describe the $\pi\pi$ system we include $\rho$, $\omega$, 
$f_2(1270)$, and three scalar mesons $f_0(600),~f_0(1370)$ and
$f_0(980)$. The masses and widths of the $\rho$,
$\omega$ and $f_2(1270)$ mesons are fixed 
at their PDG values;  the parameters of the scalar mesons 
are taken from the published papers on
the $f_0(600)$~\cite{f600}, $f_0(1370)$~\cite{f980} and
$f_0(980)$~\cite{f980} .

The contributions from the intermediate states listed above 
are included in the signal-event density ($S(q^2,q^2_1)$)
parameterization 
as a coherent sum of the 
corresponding amplitudes
together with a possible constant amplitude ($a_{ps}$). The phases of
the amplitudes are defined relative to $D_2^*$:
\begin{eqnarray}
S(q^2,q_1^2)&=&|a_{D^*_2}A^{D^*_2}(q^2,q_1^2)+a_{D^*_0}e^{i\phi_{D^*_0}}A^{D^*_0}
(q^2,q_1^2)+a_{D^*_v}e^{i\phi_{D^*_v}}A^{D^*_v}(q^2,q_1^2)\nonumber\\
&&+a_{\rho}e^{i\phi_{\rho}}(A^{\rho}(q^2,q_1^2)+r_{\omega-\rho}e^{i\phi_{\omega}}A^{\omega}(q^2,q_1^2))
+a_{f_2}e^{i(\phi_{f_2}+\phi_{\rho})}A^{f_2}(q^2,q_1^2)\nonumber\\
&&+
a_{f_0(600)}e^{i(\phi_{f_0(600)}+\phi_{\rho})}A^{f_0(600)}(q^2,q_1^2)
+a_{f_0(980)}e^{i(\phi_{f_0(980)}+\phi_{\rho})}A^{f_0(980)}(q^2,q_1^2)
\nonumber\\
&&+a_{f_0(1470)}e^{i(\phi_{f_0(1470)}+\phi_{\rho})}A^{f_0(1470)}(q^2,q_1^2)+a_{B^*}e^{i\phi_{B^*_v}}A^{B^*_v}(q^2,q_1^2)+a_{ps}e^{i\phi_{ps}}|^2,
\end{eqnarray}
where  $q^2\equiv M^2_{D\pi}$ and $q_1^2\equiv M^2_{\pi\pi}$.
The relative amplitude and phase of the 
$\omega$ meson are expressed via those of the $\rho$ meson.
The relative phase is taken from $\rho$-$\omega$ interference 
measurements~\cite{CMD2}, and the relative amplitude $(r_{\omega-\rho}=a_{\omega}/a_{\rho})$
is recalculated using 
that value. Assuming that the $\rho$ and $\omega$ mesons produced
in $B^0$ decay emerge from the 
$d\bar{d}$ pair, the relative amplitude is expected to satisfy a relation
$r_{\omega-\rho\,(B)}=-3r_{\omega-\rho\,(\gamma)}$.

We use the approach described in~\cite{mybelle}, where
each resonance is described by a relativistic Breit-Wigner function with a
$q^2$ dependent width 
and an angular dependence that corresponds to the spin and parity of the
intermediate- and final-state particles.
The $\rho$ meson amplitude is described by the
Gounaris-Sakurai parameterization~\cite{GS}.
We take into account transition form factors for hadron transitions using
the Blatt-Weisskopf parameterization~\cite{blat} with
a hadron scale $r$=1.6~$\rm(GeV/c)^{-1}$.

The variation of the detection efficiency over the Dalitz plot is taken 
into account by the minimization procedure. 
The efficiency dependence enters the likelihood function only through the
normalization term. The normalization is obtained based on a large MC
sample
generated uniformly over the Dalitz plane, processed with the same selection
criteria as the data and multiplied with the model used to fit the data.
The detector resolution for the invariant mass  of 
the $D\pi $($\pi\pi $) combination is about $2.5$ (3.5)~MeV/c$^2$,  which is
much smaller than the narrowest peak width of 30--40~MeV/c$^2$.
Hence convolution of the described parametrization with the resolution is
not necessary.
 The mass and width of the broad $(D\pi)$ resonance,   
$M_{D_0^{*+}}=2308\,\rm MeV/c^2,~\Gamma^0_{D_0^{*+}} =276\,{\rm MeV}/c^2$ 
are taken from our $D^{**0}$ measurement~\cite{mybelle}.

Table~\ref{t:dpp} gives the fit results for different models.
The contributions of different states are characterized by their
fractions, which are
defined as:
\begin{equation}
\label{e:brf}
{ f}_i= \frac{a_i^2 \int |A_i(Q)|^2 dQ}{\int |\sum_k a_ke^{i\phi_k} A_k(Q)|^2 dQ},
\end{equation} 
where $A_i(Q)$ is the corresponding amplitude, and $a_i$ and $\phi_i$ are
the amplitude coefficients and phases obtained from the fit. The
integration 
is performed over all  available phase 
space characterized by the multidimensional vector $Q$ (for decay to 3
spinless particles, $dQ\equiv dq^2dq_1^2$), and $i$ is one of the intermediate states: $D^{*}_2,~D^{*}_0,~\rho,~f_2,~f_0,~D^*_v,~B^*_v$ or 
the constant term $a_{ps}$. The sum of
the individual fractions ${f_i}$ exceeds  unity for our case
because of destructive interference.
The product of the branching fractions of the $B$ meson is expressed
via the 
fraction ${f}_i$:
\begin{equation}
\label{e:brf2}
{\cal B}_{B\to i\pi}{\cal B}_{i\to D\pi}= \frac{N_{\rm sig}{f}_i}{N_{B\bar{B}}},
\end{equation} 
where $N_{\rm sig}$ is the efficiency corrected number of the
 reconstructed $D\pi\pi$ events and
 $N_{B\bar{B}}$ is the number of  ${B\bar{B}}$ pairs produced.
\begin{table}
 \begin{tabular}{|c|c|c|c|c|c|}
 \hline
Options & 1 & 2
 &3       & 4      & 5       \\
 \hline
 States& $D^*_2,~D^*_0,~D^*_v,$ &
 $D^*_2,~D^*_0,$
 &$D^*_2,~D^*_0,~D^*_v,$       & $D^*_2,~D^*_0,~D^*_v,$       & $D^*_2,~D^*_v,$       \\
 & $\rho,~f_2,~f_0's$ &
 $\rho,~f_2,~f_0's$
 &$\rho,~f_2,~f_0's,~B^*_v$       & $\rho,~f_2,~f_0's+ps$       & $\rho,~f_2,~f_0's$       \\
 \hline
  $ -2\ln{\cal L}/{\cal L}_0$ &   0    &   69.5    &   -2.7    &  -13.0     &  51.3    \\
 \hline
$N_1$                      & $    2181\pm      64$& $    2174\pm      64$& 	 $    2223\pm      71$&	 $    2264\pm      65$&	 $    2111\pm      62$\\
 \hline			      			     			     			  			  			 
${\cal B}_{B\to D_2^*\pi}{\cal B}_{D_2^*\to D\pi}(10^{-4})$      & $    2.15\pm    0.16$& $    2.23\pm    0.12$& 	 $    2.15\pm    0.18$&	 $    2.26\pm    0.18$&	 $    2.51\pm    0.14$\\
$M_{D^*_2},(MeV/c^2)$      & $  2465.7\pm     1.7$& $  2461.9\pm     1.6$& 	 $  2465.2\pm     1.9$&	 $  2464.9\pm     1.6$&	 $  2465.5\pm     1.7$\\
$\Gamma_{D^*_2},(MeV/c^2)$ & $    49.6\pm     3.9$& $    49.0\pm     3.9$& 	 $    49.3\pm     4.1$&	 $    51.5\pm     3.8$&	 $    55.4\pm     4.0$\\
 \hline			       			      			      			   			   			 
${\cal B}_{B\to D_0^*\pi}{\cal B}_{D_0^*\to D\pi}(10^{-4})$      & $    0.60\pm    0.13$& $    0.61\pm    0.10$& 	 $    0.50\pm    0.13$&	 $    0.79\pm    0.11$&	 $    0$\\
$\phi_{D_0^*}$             & $   -3.00\pm    0.13$& $   -2.28\pm    0.17$& 	 $   -2.88\pm    0.17$&	 $   -2.66\pm    0.11$&	 $   0$\\
$M_{D^*_0},(MeV/c^2)$      & $  2308.0$&	    $  2308.0$&		  $2308.0$&		 $  2308.0$&		 $  2308.0$\\
$\Gamma_{D^*_0},(MeV/c^2)$ & $   276.0$&	    $   276.0$&		  $276.0$&		 $   276.0$&		 $   276.0$\\
 \hline			       			      			      			   			   			 
${\cal B}_{B\to D_v^*\pi}{\cal B}_{D_v^*\to D\pi}(10^{-4})$      & $    0.88\pm    0.13$& $    0$& 	 $    0.85\pm    0.14$&	 $    0.74\pm    0.11$&	 $    0.66\pm    0.10$\\
$\phi_{D^*_v}$             & $   -2.62\pm    0.15$& $   0$&		   $   -2.53\pm    0.17$&	 $   -2.59\pm    0.13$&	 $   -3.04\pm    0.20$\\
 \hline			       			      			      			   			   			 
${\cal B}_{B\to D\rho}{\cal B}_{\rho\to \pi\pi}(10^{-4})$       & $    3.19\pm    0.20$& $    2.94\pm    0.15$& 	 $    3.15\pm    0.21$&	 $    3.07\pm    0.14$&	 $    3.26\pm    0.18$\\
$\phi_{\rho}$              & $    2.25\pm    0.19$& $    1.45\pm    0.22$& 	 $    1.81\pm    0.31$&	 $    1.73\pm    0.15$&	 $    1.89\pm    0.15$\\
$M_{\rho},(MeV/c^2)$       & $   775.6$&	    $   775.6$&		  $775.6$&		 $   775.6$&		 $   775.6$\\
$\Gamma_{\rho},(MeV/c^2)$  & $   144.0$&	    $   144.0$&		  $144.0$&		 $   144.0$&		 $   144.0$\\
 \hline			       			      			      			   			   			 
${\cal B}_{B\to Df_2}{\cal B}_{f_2\to \pi\pi}(10^{-4})$        & $    0.68\pm    0.10$& $    0.64\pm    0.08$& 	 $    0.64\pm    0.09$&	 $    0.54\pm    0.08$&	 $    0.65\pm    0.08$\\
$\phi_{f_2}$               & $    2.97\pm    0.21$& $    2.48\pm    0.16$& 	 $    2.77\pm    0.20$&	 $    2.32\pm    0.13$&	 $    2.91\pm    0.17$\\
$M_{f_2},(MeV/c^2)$        & $  1275.0$&	    $  1275.0$&		  $1275.0$&		 $  1275.0$&		 $  1275.0$\\
$\Gamma_{f_2},(MeV/c^2)$   & $   185.0$&	    $   185.0$&		  $185.0$&		 $   185.0$&		 $   185.0$\\
 \hline			       			      			      			   			   			 
${\cal B}_{B\to Df_0(600)}{\cal B}_{f_0(600)\to \pi\pi}(10^{-4})$   & $    0.68\pm    0.08$& $    0.72\pm    0.09$& 	 $    0.72\pm    0.09$&	 $    0.47\pm    0.08$&	 $    0.58\pm    0.07$\\
$\phi_{f_0(600)}$          & $   -0.44\pm    0.09$& $   -0.42\pm    0.09$& 	 $   -0.40\pm    0.10$&	 $   -0.43\pm    0.11$&	 $   -0.32\pm    0.10$\\
$M_{f_0},(MeV/c^2)$        & $   513.0$&	    $   513.0$&		  $513.0$&		 $   513.0$&		 $   513.0$\\
$\Gamma_{f_0},(MeV/c^2)$   & $   335.0$&	    $   335.0$&		  $335.0$&		 $   335.0$&		 $   335.0$\\
 \hline			       			      			      			   			   			 
${\cal B}_{B\to Df_0(980)}{\cal B}_{f_0(980)\to \pi\pi}(10^{-4})$   & $    0.08\pm    0.04$& $    0.11\pm    0.04$& 	 $    0.08\pm    0.04$&	 $    0.04\pm    0.02$&	 $    0.08\pm    0.03$\\
$\phi_{f_0(980)}$          & $   -2.48\pm    0.47$& $    2.68\pm    0.37$& 	 $   -3.07\pm    0.51$&	 $    2.87\pm    0.37$&	 $   -2.85\pm    0.32$\\
$M_{f_0},(MeV/c^2)$        & $   978.0$&	    $   978.0$&		  $978.0$&		 $   978.0$&		 $   978.0$\\
$\Gamma_{f_0},(MeV/c^2)$   & $    44.0$&	    $    44.0$&		  $44.0$&		 $    44.0$&		 $    44.0$\\
 \hline			       			      			      			   			   			 
${\cal B}_{B\to Df_0(1370)}{\cal B}_{f_0(1370)\to \pi\pi}(10^{-4})$  & $    0.21\pm    0.10$& $    0.24\pm    0.06$& 	 $    0.18\pm    0.06$&	 $    0.15\pm    0.04$&	 $    0.25\pm    0.10$\\
$\phi_{f_0(1370)}$         & $   -1.52\pm    0.56$& $    3.08\pm    0.35$& 	 $   -2.43\pm    0.62$&	 $   -2.75\pm    0.28$&	 $   -2.00\pm    0.38$\\
$M_{f_0},(MeV/c^2)$        & $  1434.0$&	    $  1434.0$&		  $1434.0$&		 $  1434.0$&		 $  1434.0$\\
$\Gamma_{f_0},(MeV/c^2)$   & $   173.0$&            $   173.0$&             $   173.0$&             $   173.0$&             $   173.0$\\
 \hline
${\cal B}_{B\to B^*_v\pi}{\cal B}_{B^*_v\to D\pi}(10^{-4})$      & 0 &0 &$0.74\pm0.76$ &0 &0 \\
$\phi_{B^*_v}$             & 0 &0 &$1.09\pm0.51$ &0 &0 \\
 \hline				      	 															
$\phi_{ps}$                & 0 &0 &0 &$0.22\pm0.14$ &0 \\
${\cal B}_{ps}(10^{-4})$         & 0 &0 &0 &$0.33\pm0.18$ &0 \\
\hline
$\chi^2/N_{dof}$& 629/603&680/605&632/601&618/601&659/605\\ 
$CL~(\%)$& 23&1.8&18&31&6.3\\
\hline
 \end{tabular}
\caption{The fit results for different sets of amplitudes. Option
  1 is used as the main set of amplitudes for the final results.  }
\label{t:dpp}
\end{table}

Table~\ref{t:dpp} contains information on the likelihood change
relative to the main set, and $\chi^2$ values obtained from four histograms:
projections of $M_{D\pi}$ and $M_{\pi\pi}$ for negative and positive
helicities of the $D\pi$ and $\pi\pi$ systems,
respectively\cite{footchi}.

The fit gives a  statistically significant  contribution
from off-shell $D^*_v\pi$ production; the addition
of the off-shell $B^*_v$ amplitude does not improve
the likelihood value significantly. The inclusion of the
three-particle 
phase space term 
improves the likelihood value, however there is no  reason  to expect 
a constant amplitude with no momentum dependence over such a wide range
 of final particle momenta.  
Table~\ref{t:jp} shows that
the likelihood changes significantly when 
the broad resonance $D^*_0$ is removed or treated as 
either a vector or a
tensor. The change of likelihood $-2\ln{\cal L}/{\cal L}_0=51$ for 2
additional degrees of freedom (amplitude and phase of $D^*_0$) 
corresponds to a significance of
6.8~$\sigma$~\cite{wilk}.

The branching fractions of $D^*_2$
and $D_0^*$ remain constant within errors for different models. 
 The set of 
states  used for the  final results are  
$D^*_2,~D^*_0,~D^*_v,~\rho,~f_2$ and the
three above-listed $f_0$'s listed above, corresponding to column~1 in
Table~\ref{t:dpp}.

The values of the $D_2^{*+}$ resonance mass and width obtained from
the fit are:
$$
 M_{D^{*+}_2}=(2465.7\pm1.8\pm0.8^{+1.2}_{-4.7}) {\,\rm MeV}/c^2,~~\Gamma_{D^{*+}_2}=(49.7\pm3.8\pm4.1\pm4.9){\,\rm MeV},
$$
where the third error is model uncertainty.
These parameters are consistent with but more precise than previous 
 measurements performed by 
CLEO $
M_{D^{*0}_2}=(2463\pm3\pm3)\, {\rm MeV}/c^2
$~\cite{dobs} and FOCUS 
$
~\Gamma_{D^{*}_2}=(34.1\pm6.5\pm4.2)\, \rm{MeV}
$~\cite{FOC}.

The product of the branching fraction for $D^*_2$ production obtained from the
fit is:
$$
{\cal B}(\bar{B}^0\to D^{*+}_2\pi^-)\times {\cal B}(D_2^{*+}\to D^{0}\pi^+)=(2.15\pm0.17\pm0.29\pm0.12)\times10^{-4},
$$
where the three errors are statistical, systematic,
and a model-dependent error, respectively.
We observe the production of the broad scalar $D_0^{*+}$ state with
the product  
branching fraction,
$$
{\cal B}(\bar{B}^0\to D^{*+}_0\pi^-)\times{\cal B}(D_0^{*+}\to D^{0}\pi^+)=(0.60\pm0.13\pm0.15\pm0.22)\times10^{-4}.
$$
This is the first observation of this decay (the interpretation of the
neutral partner of this state is still a subject of theoretical
discussion~\cite{tfoc}). 
The relative phase of the $D_0^*$ amplitude is
$$
\phi_0=3.00\pm0.13\pm0.10\pm0.43.
$$

\begin{figure}[h]
\begin{center}
\begin{tabular}[c]{cc}
\includegraphics[height=8 cm]{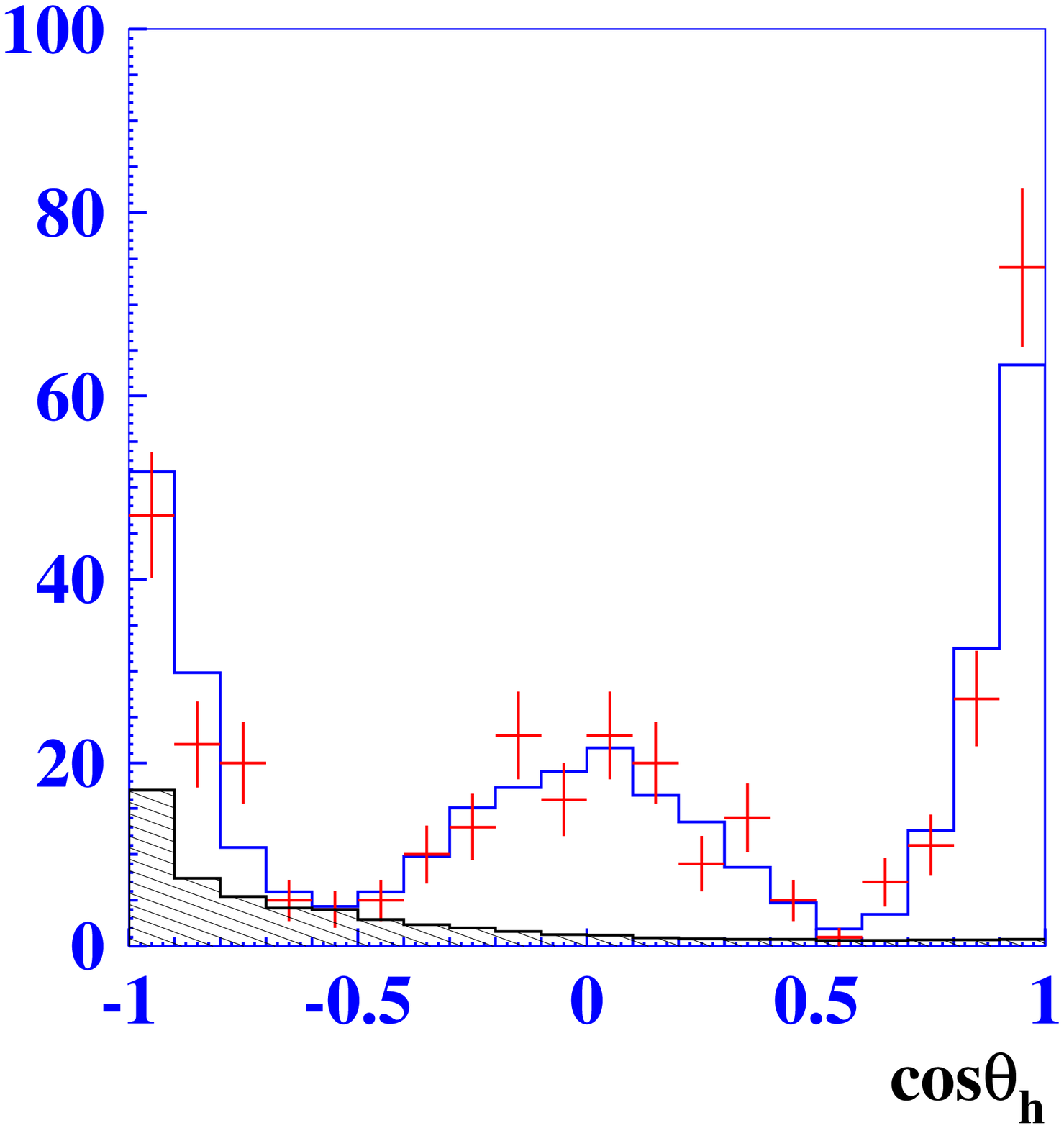}&
\includegraphics[height=8 cm]{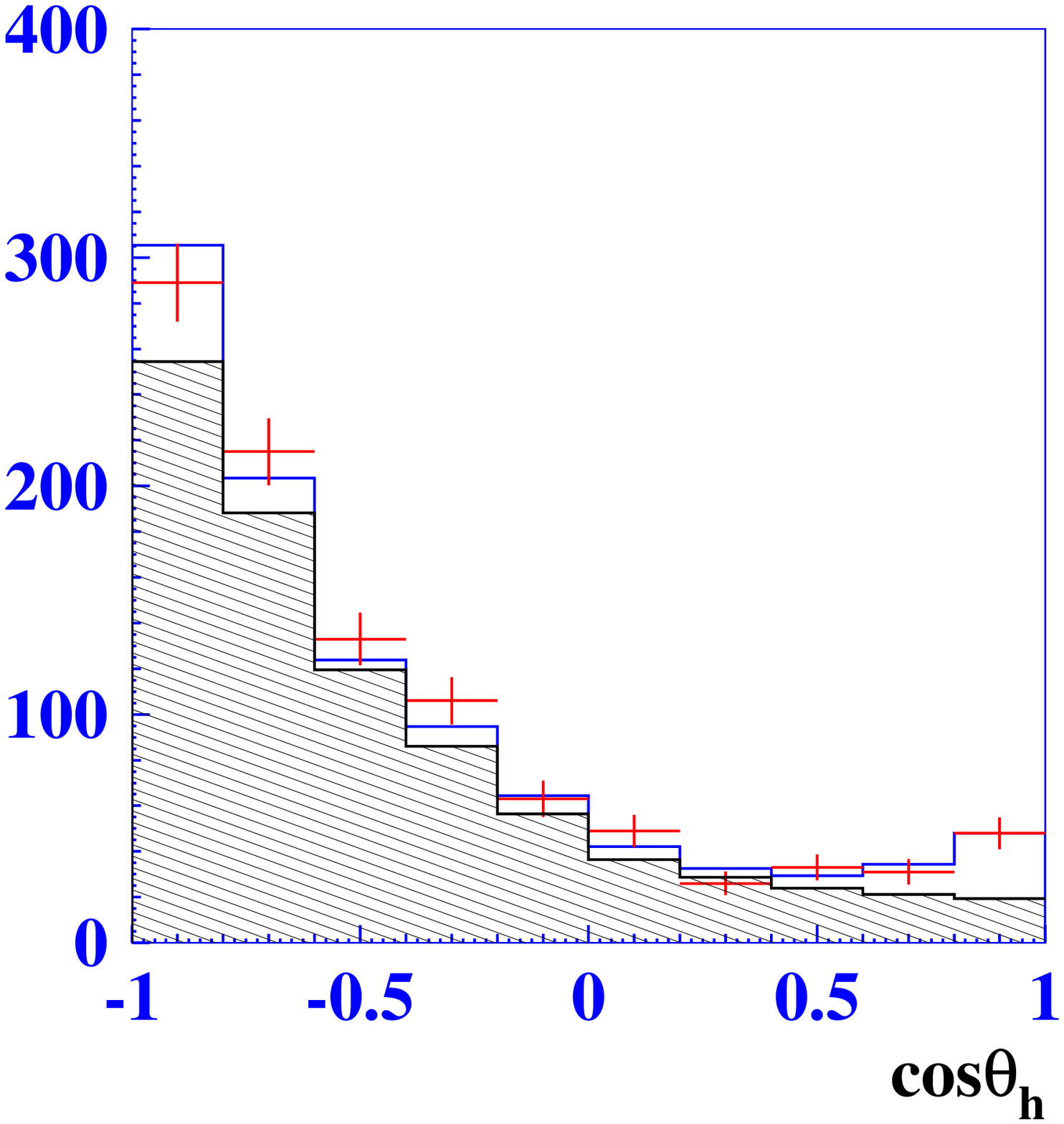}\\
\vspace*{-8 cm} & \\
{\bf\large \hspace*{-3cm} a)}&{\bf\large \hspace*{-3cm} b)}\\
\vspace*{6.5 cm} & \\
\end{tabular}
\end{center}
\caption{$D\pi$ helicity angle distributions for 
data (points) 
and MC (histogram). The hatched distribution shows 
the background from the $\Delta E$ sideband region with the
appropriate normalization. (a) corresponds to the $D^*_2$ region
$|M_{D\pi}-2.46|<0.1\rm\,GeV/c^2$; (b) the
$D_0$ region $|M_{D\pi}-2.30|<0.1\rm\,GeV/c^2$.
} 
\label{f:dpp_hel}
\end{figure}

\noindent
The $D\pi$ helicity angle
distributions for  $ M_{D\pi}$ regions corresponding to
the $D^*_2$ and $D^*_0$ are shown in
Fig.~\ref{f:dpp_hel}(a) and (b), respectively, 
together with 
the efficiency-corrected fitting function.
The histogram 
in the region of the $D^*_2$ meson clearly indicates a D-wave. The 
distributions in the other  regions 
show  reasonable agreement between the fitting function 
and the data.  

\begin{table}
 \begin{tabular}{|c|c|c|c|c|}
 \hline
& no broad state
 & $0^+$       & $1^-$      & $2^+$       \\
 \hline
  $ -2\ln{\cal L}/{\cal L}_0$ &  51    &   0& 28    &   27    \\
 \hline
$\chi^2/N_{dof}$&659/605& 629/603&652/603&640/603\\ 
$CL,~(\%)$&6.3 & 22&8.1&14\\
\hline
 \end{tabular}
\caption{Changes in likelihood for different quantum number
assignments for the broad resonance.}
\label{t:jp}
\end{table}

The uncertainty of the background is one of the main sources of systematic error.
This is estimated by comparing the fit results for 
the case when the  background shape is taken 
separately from the lower or upper 
$\Delta E$ sidebands. The fit is also performed 
with more 
restrictive and looser cuts on $\Delta E$, $M_{\rm bc}$ and $\Delta M_D$
that change the signal-to-noise ratio by factors of about two.
The results obtained are consistent with each other and 
the maximum difference is taken as an additional estimate 
of the systematic uncertainty.   
The
systematic errors on the ${\cal B}_i$ measurements (Eq.~(\ref{e:brf2})) 
for the individual intermediate
states include 
uncertainties in  track reconstruction and PID efficiency,
as well as the error in the $D^0\to K^-\pi^+$ absolute 
branching fraction.
The model uncertainties are estimated by comparing fit results 
for the case of different models
and for values of  the parameter $r$ of the  transition form
factor~\cite{blat}
from 0 to 3 (GeV/$c$)$^{-1}$. 

\begin{figure}[h]
\begin{center}
\begin{tabular}[c]{ccc}
\includegraphics[height=5 cm]{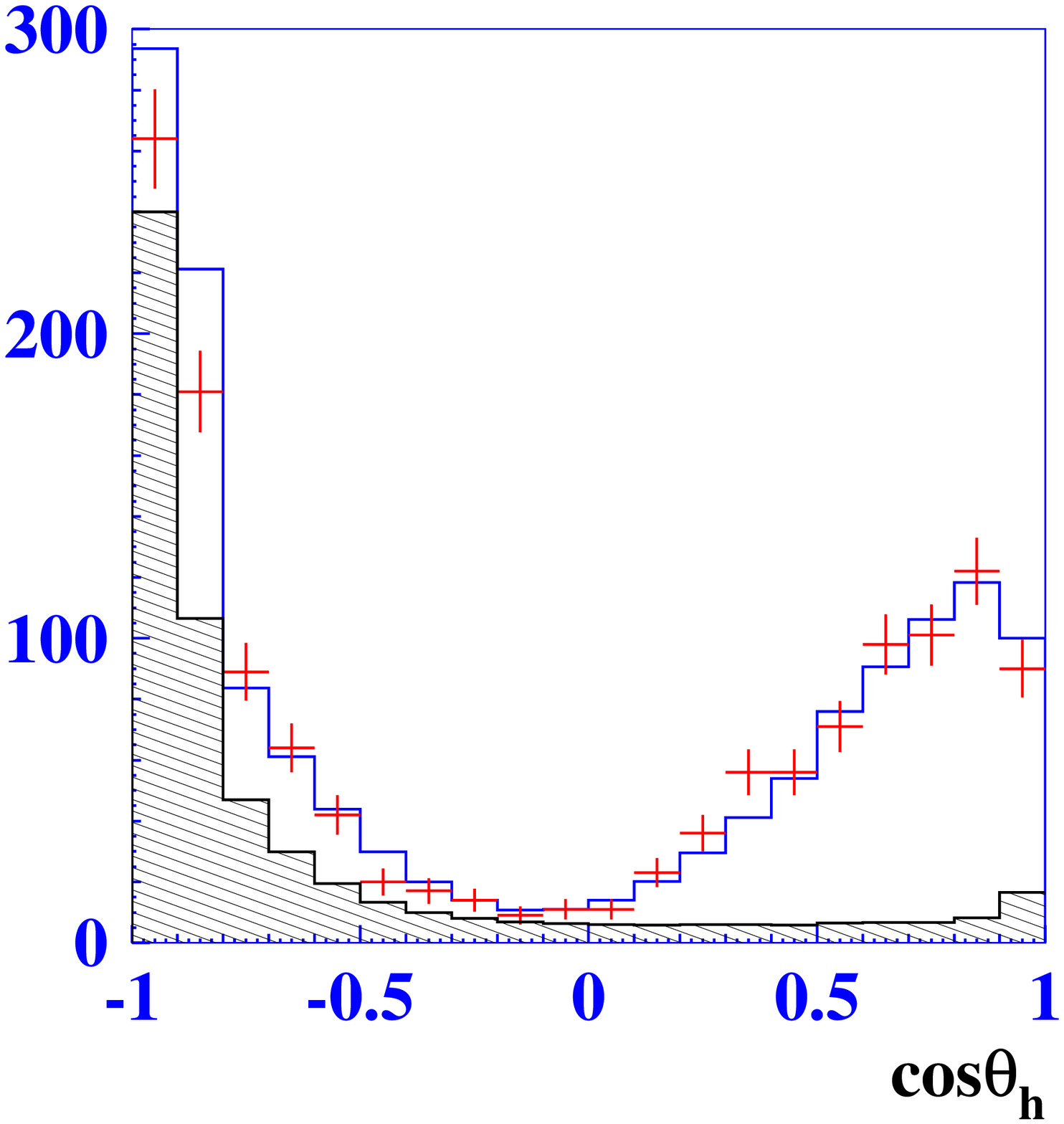}&
\includegraphics[height=5 cm]{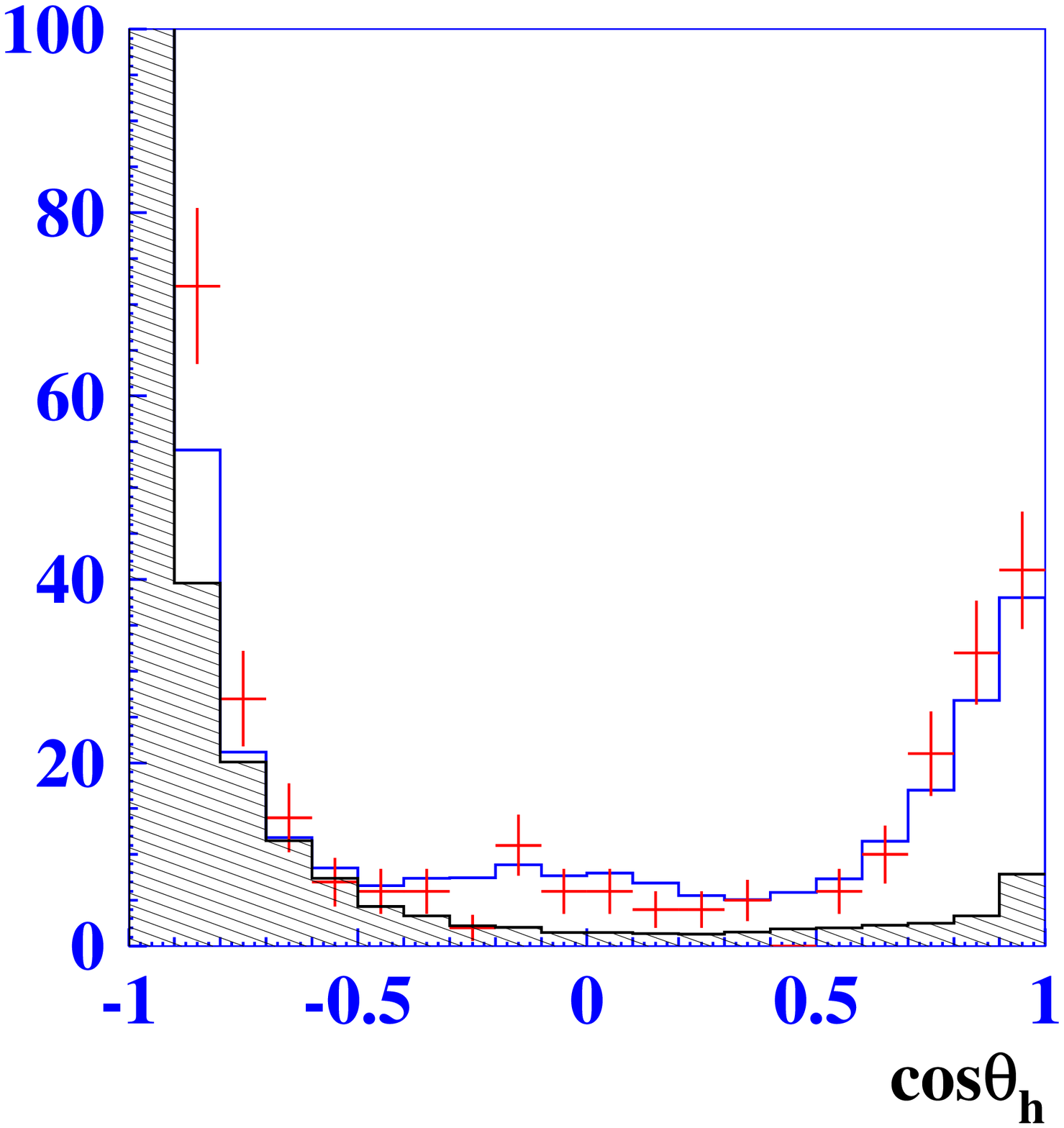}&
\includegraphics[height=5 cm]{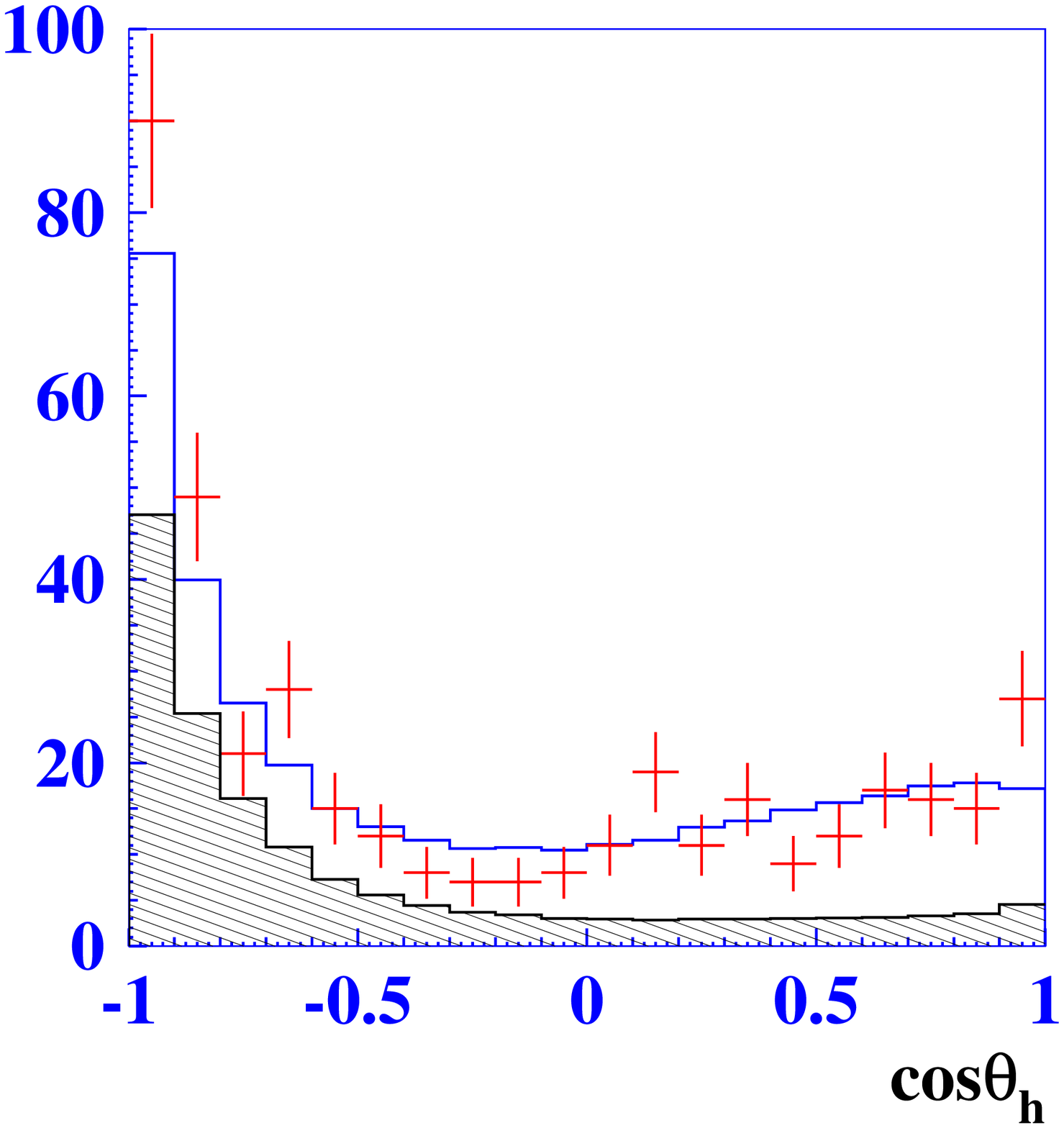}\\
\vspace*{-5 cm} & \\
{\bf\large \hspace*{-1cm} a)}&{\bf\large \hspace*{-1cm} b)}&{\bf\large \hspace*{-1cm} c)}\\
\vspace*{3 cm} & \\
\end{tabular}
\end{center}
\caption{$\pi\pi$ helicity angle distributions for data (points) and 
MC (histogram). The hatched distribution shows 
the background distribution from the $\Delta E$ sideband region with appropriate 
normalization. (a) corresponds to the $\rho$ region $|M_{\pi\pi}-0.78|<0.2
\,$GeV/$c^2$; (b) the $f_2$ region $|M_{\pi\pi}-1.20|<0.1\,$GeV/$c^2$; (c) the $f_0$ region 
$M_{\pi\pi}<0.60\,$GeV/$c^2$.}
\label{f:pp_hel}
\end{figure}

The $\pi\pi$ helicity angle distributions for $M_{\pi\pi}$ ranges
corresponding to the
$\rho$, $f_2$ and the region below the $\rho$, 
where the broad resonance dominates,
are shown in Fig.~\ref{f:pp_hel}.
For the positive helicity region, where the $D\pi$ contribution is
suppressed, a clear $P$-wave structure for the $\rho$ and 
$D$-wave structure for the $f_2$ is observed.
The scalar component parameters cannot be determined from 
the fit. This process can also have contributions from non-resonant 
background. 
The product branching fraction for the $f_2$ is 
$$
{\cal B}(\bar{B}^0\to  D^0 f_2){\cal B}(f_2\to \pi^+\pi^-)=
(0.68\pm0.10\pm0.12\pm0.18)\times10^{-4}.
$$
Taking into account the
branching fraction ${\cal B}(f_2\to \pi\pi)=$
$0.847^{+0.025}_{-0.012}$~\cite{PDG} and the corresponding Clebsch-Gordan
 coefficients, we obtain
$$
{\cal B}(\bar{B}^0\to  D^0 f_2)=
(1.20\pm0.18\pm0.21\pm0.32)\times10^{-4},
$$
$$
{\cal B}(\bar{B}^0\to  D^0 \rho^0)=
(3.19\pm0.20\pm{0.24}\pm0.38)\times10^{-4}.
$$
The phases relative to the $D_2^*$ amplitude are $
\phi_{\rho}=2.25\pm0.19\pm0.20^{+0.21}_{-0.99}$ and $
~\phi_{f_2}=2.97\pm0.21\pm0.13\pm{0.45}.
$

\subsection{Results and discussion}

The branching fraction products obtained for the narrow $(j=3/2)$
resonances are similar to
the published results for charged $B$
decays as shown in Table~\ref{zak}.
\begin{table}
\begin{tabular}{|c|c|c|}
\hline
& neutral $B$ & charged $B$~\cite{mybelle}\\
\hline
$
{\cal B}(\bar{B}\to D^{*}_2\pi^-){\cal B}(D_2^{*}\to
D\pi)
$
&
$
(2.15\pm0.17\pm0.29\pm0.12)\times10^{-4}
$
&
$
(3.4\pm0.3\pm0.6\pm0.4)\times10^{-4}
$\\
$
{\cal B}(\bar{B}\to D^{*}_2\pi^-){\cal B}(D_2^{*}\to
D^{*}\pi)
$
&
$
(2.45\pm0.42^{+0.35+0.39}_{-0.45-0.17})\times10^{-4}
$~\cite{hhh}
&
$
(1.8\pm0.3\pm0.3\pm0.2)\times10^{-4}
$\\
$
{\cal B}(\bar{B}\to D_1\pi^-){\cal B}(D_1\to
D^{*}\pi)
$
&
$
(3.68\pm0.60^{+0.71+0.65}_{-0.40-0.30})\times10^{-4}
$~\cite{hhh}
&
$
(6.8\pm0.7\pm1.3\pm0.3)\times10^{-4}
$\\
\hline
$
{\cal B}(\bar{B}\to D^{*}_0\pi){\cal B}(D_0^{*}\to D\pi)
$
&
$
(0.60\pm0.13\pm0.15\pm0.22)\times10^{-4}
$
&
$
(6.1\pm0.6\pm0.9\pm1.6)\times10^{-4}
$\\
$
{\cal B}(\bar{B}\to D'_1\pi^-){\cal B}(D'_1\to
D^{*}\pi)
$
&
$
<0.7\times10^{-4} \rm ~at~90\,\%~C.L.
$~\cite{hhh}
&
$
(5.0\pm0.4\pm1.0\pm0.4)\times10^{-4}
$\\
\hline
\end{tabular}
\caption{Comparison of product branching fractions for neutral
  and charged $B$ decays.}
\label{zak}
\end{table}
The measured values of the branching fractions for the broad 
$D^{*+}_0$  resonances in neutral $B$ decays are, however, 
significantly lower than those for charged $B$ decays.
Preliminary data on $\bar{B^0}\to D^{*0}\pi^+\pi^-$ decay~\cite{hhh}
shown in Table~\ref{zak} indicates
a similar behavior for  $D'^{0}_1$ and $D_1^{0}$ production.
One possible explanation for this phenomenon is that 
for charged $B$ decay to $D^{**}\pi$, the amplitude
receives contributions from both tree and 
color-suppressed diagrams as shown in Fig.~\ref{f:fd1}.
For the color-suppressed diagrams, however, $D^{**}$'s are produced by
another 
mechanism and the amplitudes are characterized by the 
constants $f_{D(3/2)}$ and 
$f_{D(1/2)}$, with $f_{D(3/2)} \ll f_{D(1/2)}$.
The production of the broad resonances 
$D^{*0}_0$ and $D'^{0}_1$ in charged $B$ decay is 
amplified by
the color-suppressed amplitude. As shown in~\cite{yon2}, in such a
case both $\tau_{1/2}$ fits to the sum rule and the value of
$f_{D(1/2)}$ are consistent with theoretical estimates.

\section{Conclusion}

A study of neutral $B$-meson decays to $D^0\pi^+\pi^-$ 
is reported.  
We measure the total branching fraction of the three-body
$D^0\pi^+\pi^-$  decays, obtaining
${\cal B}(\bar{B}^0\to D^0\pi^+\pi^-)
=(8.4\pm0.4\pm0.8)\times10^{-4}$.
The intermediate resonant structure  of these three-body decays is studied.
The $D^0\pi^+\pi^-$ final state is described by the  
production of $D^*_{0,2}\pi^-$  with
subsequent decays $D^{*}_{0,2}\to D\pi$,  and also by $D\rho,~Df_2$,
and $D\sigma$, where $\sigma$  is a 
broad scalar ($\pi\pi$) structure.
From a Dalitz plot analysis we obtain the mass, width and  
product of the branching fractions for the $D^{*+}_2$: 
$$
M_{D^{*+}_2}=(2465.7\pm1.8\pm0.8^{+1.2}_{-4.7}) {\rm MeV}/c^2,~~\Gamma_{D^{*+}_2}=(49.7\pm3.8\pm4.1\pm4.9){\rm MeV},
$$
$$
{\cal B}(\bar{B}^0\to D^{*+}_2\pi^-)\times {\cal B}(D_2^{*+}\to D^{0}\pi^+)=
(2.15\pm0.17\pm0.29^\pm0.12)\times10^{-4}.
$$
\noindent
We observe the production of the broad scalar $D_0^{*+}$ state   with 
the product branching fraction
$$
{\cal B}(\bar{B}^0\to D^{*+}_0\pi^-)\times {\cal B}(D_0^{*+}\to
D^{0}\pi^+)=(0.60\pm0.13\pm0.15\pm0.22)\times10^{-4}.
$$
This is the first observation of this decay. 
The phase of the $D_0^*$ amplitude relative  to that of the  $D_2^*$ is determined to be:
$$
\phi_0=3.00\pm0.13\pm0.10\pm0.43.
$$

The $B \to D\rho$ and $Df_2$ branching fractions are 
measured to be:
$$
{\cal B}(\bar{B}^0\to  D^0 \rho^0)=
(3.19\pm0.20\pm{0.24}\pm0.38)\times10^{-4},
$$
$$
{\cal B}(\bar{B}^0\to  D^0 f_2)=
(1.20\pm0.18\pm0.21\pm0.32)\times10^{-4},
$$
and the phases relative to the $D_2^*$ amplitude are:
$$
\phi_{\rho}=2.25\pm0.19\pm0.20^{+0.21}_{-0.99},
$$
$$
\phi_{f_2}=2.97\pm0.21\pm0.13\pm{0.45}.
$$
This is the first observation of the $\bar{B^0}\to D^0 f_2$ decay.

\section*{Acknowledgments}
We thank the KEKB group for the excellent operation of the
accelerator, the KEK Cryogenics group for the efficient
operation of the solenoid, and the KEK computer group and
the National Institute of Informatics for valuable computing
and Super-SINET network support. We acknowledge support from
the Ministry of Education, Culture, Sports, Science, and
Technology of Japan and the Japan Society for the Promotion
of Science; the Australian Research Council and the
Australian Department of Education, Science and Training;
the National Science Foundation of China under contract
No.~10175071; the Department of Science and Technology of
India; the BK21 program of the Ministry of Education of
Korea and the CHEP SRC program of the Korea Science and
Engineering Foundation; the Polish State Committee for
Scientific Research under contract No.~2P03B 01324; the
Ministry of Science and Technology of the Russian
Federation; the Ministry of Education, Science and Sport of
the Republic of Slovenia; the National Science Council and
the Ministry of Education of Taiwan; and the U.S.\
Department of Energy.


\begin{thebibliography}{120}
\bibitem{foot1}{The $z$ coordinate of the track is
defined as the $z$ coordinate of the track point closest to the beam in
the $r-\phi$ 
plane. The $z$  axis is opposite to the positron beam direction.}
\bibitem{footchi}{The overall $\chi^2$ is calculated as the sum of the
$\chi^2$'s from four distributions:  two 160-bin histograms of $M_{D\pi}$
for positive
and negative helicities of the $D\pi$ system, and two 150-bin histograms
$M_{\pi\pi}$
 with positive
and negative helicities of the $\pi\pi$ system. The number of degrees of
freedom is calculated as the number of bins minus the number of free parameters.}
\end{thebibliography}
\end{document}